\pdfoutput=1
\documentclass{JINST}
\let\ifpdf\relax

\newcommand{\cskc}{~s$^{-1}$keV$^{-1}$cm$^{-2}$}
\newcommand{\ckcs}{~keV$^{-1}$cm$^{-2}$s$^{-1}$}

\title{Low background x-ray detection with Micromegas for axion research}

\author{S.~Aune$^a$, J.F.~Castel$^b$ T.~Dafni$^b$, M.~Davenport$^e$, G.~Fanourakis$^c$, E.~Ferrer-Ribas$^a$, J.~Gal\'an$^b$, J.A.~Garc\'ia$^b$, A.~Gardikiotis$^d$, T.~Geralis$^d$, I.~Giomataris$^a$, H.~G\'omez$^b$, J.G.~Garza$^b$, D.C.~Herrera$^b$, F.J.~Iguaz$^{a,b}$, I.G.~Irastorza$^b$, D.~Jourde$^a$ , G.~Luz\'on$^b$, J.P.~Mols$^a$, T.~Papaevangelou$^a$, A.~Rodr\'iguez$^b$, J.~Ruz$^{e,f}$, L.~Segu\'i$^{b, } $\footnote{Present addr.: University of Oxford, Oxford, UK } , A. Tom\'as$^{b, }$\footnote{Corresponding author}  \footnote{Present addr.: Brackett Laboratory, Imperial College, London, UK}, T.~Vafeiadis$^{e,h}$, S.C.~Yildiz$^{i,j}$  \\
\llap{$^a$} Centre d'\'Etudes de Saclay, CEA, Gif-sur-Yvette, France\\
\llap{$^b$} Grupo de F\'isica Nuclear y Astropart\'iculas, University of Zaragoza, Zaragoza, Spain\\
\llap{$^c$} Institute of Nuclear Physics, NCSR Demokritos, Athens, Greece\\
\llap{$^d$} University of Patras, Patras, Greece\\
\llap{$^e$} CERN, European Organization for Particle Physics and Nuclear Research, Geneva, Switzerland\\
\llap{$^f$} Lawrence Livermore National Laboratory, Livermore, CA, USA\\
\llap{$^h$} Aristotle University of Thessaloniki Thessaloniki, Greece\\
\llap{$^i$} Do\u{g}u\c{s} University, Istanbul, Turkey\\
\llap{$^j$} Bo\u{g}azi\c{c}i University,  Istanbul, Turkey\\
  E-mail: \email{a.tomas-alquezar@imperial.ac.uk}}

\abstract{
Axion helioscopes aim at the detection of solar axions through their conversion into x-rays in laboratory magnetic fields. The use of low background x-ray detectors is an essential component contributing to the sensitivity of these searches. Here we review the recent advances on Micromegas detectors used in the CERN Axion Solar Telescope (CAST) and proposed for the future International Axion Observatory (IAXO). The most recent Micromegas setups in CAST have achieved background levels of 1.5$\times$10$^{-6}$\ckcs, a factor of more than 100 lower than the ones obtained by the first generation of CAST detectors. This improvement is due to the development of active and passive shielding techniques, offline discrimination techniques allowed by highly granular readout patterns, as well as the use of radiopure detector components. The status of the intensive R\&D to reduce the background levels will be described, including the operation of replica detectors in test benches and the detailed Geant4 simulation of the detector setup and the detector response, which has allowed the progressive understanding of background origins. The best levels currently achieved in a test setup operating in the Canfranc Underground Laboratory (LSC) are as low as $\sim$10$^{-7}$\ckcs, showing the good prospects of this technology for application in the future IAXO.}


\begin{document}

\section{Introduction}

The existence of axions is motivated by the Peccei-Quinn (PQ) mechanism for solving the long-standing strong-CP problem of QCD ~\cite{Peccei:1977hh}, one of the most serious shortcomings of the standard model (SM)~\cite{Cheng:1987gp}. Similar axion-like particles (ALP) or more general Weakly Interacting Slim Particles (WISPs) appear in a number of extensions of the SM and in particular in string theory~\cite{Ringwald:2012cu}. More importantly, axions (and ALP) are also very compelling candidates to be the dark matter ~\cite{Abbott:1982af,Dine:1982ah,Preskill:1982cy}. The search for WISPs at the low-energy high-intensity frontier is increasingly recognized as a relevant portal for new physics beyond the SM, complementary to the high energy frontier at accelerators.


\medskip
Most searches for axions and ALPs rely on their conversion into photons in electromagnetic fields~\cite{Sikivie:1983ip}. Although other detection phenomenology is possible (e.g. axion-electron interaction), that conversion is guaranteed thanks to the generic two-photon vertex. Axion helioscopes are an important category of experimental searches, making use of the Sun as a hypothetical powerful source of axions. The CERN Axion Solar Telescope (CAST) ~\cite{Zioutas:2004hi} is currently the most powerful implementation of an axion helioscope. The main element of a helioscope is the use of a powerful long magnet able to point and track the Sun. The coherence over a large magnetic distance overcomes the smallness of the axion interaction strength, allowing for detectable conversion probabilities. The CAST magnet (a LHC decommissioned superconducting test magnet) enjoys a maximum 9~T field over 10~m of length.

\medskip
Solar axions are in the range of 1--10 keV, and so are the converted photons. One of the experimental challenges of axion helioscopes is thus the detection of as low an x-ray flux as possible. For this, the use of low background x-ray detection techniques is needed, potentially coupled to x-ray optics to focalize the parallel beam of photons into a small spot, in order to further increase the signal-to-noise ratio. CAST is currently using four x-ray detection systems, attached to each of the magnet bore exits (the CAST magnet is twin apperture). Two of them are in the sunset magnet side, and the other two at the sunrise one, the naming coming from the time of the day they are in axion-sensitive conditions (i.e., the magnet pointing to the Sun). Three of them (two at the sunset side, and the other at the sunrise one) follow a similar configuration, based on a shielded small gaseous chamber with a \textit{microbulk} Micromegas pixelated readout, with a thin x-ray-transparent window facing the $\sim$15~cm$^{2}$ magnet bore area. 
The fourth system is a CCD coupled to an x-ray optics (a spare one from the ABRIXAS x-ray astronomy mission~\cite{kuster2007}) able to focus the photons down to a few mm diameter spot. Both kinds of system achieve low effective backgrounds by different strategies (the first one by means of low detector backgrounds, and the second one by signal focalization). Future plans aim at combining both strengths into the same system. For the forthcoming data taking campaign, a new x-ray optics will be installed in one of the current Micromegas lines~\cite{ThPapaevangelou:SPSC2012} (and the current CCD will also be replaced by an Ingrid detector, a Micromegas-like detector with a different readout~\cite{unser_cast_paper,proc_ingrid}). In this paper we focus on the low background Micromegas detectors.

\medskip
So far CAST has achieved leading results in the search for axions ~\cite{Andriamonje:2007ew,Arik:2008mq,Aune:2011rx,Arik:2013nya}, providing a bound on the axion photon coupling at the level of 10$^{-10}$ GeV$^{-1}$ for axion masses up to $\sim$1 eV, the more stringent experimental bound in most of this axion mass range. Better results are expected after the detector improvements previously mentioned. However, a very motivated large step in sensitivity could be achieved with a new generation axion helioscope like the recently proposed International Axion Observatory (IAXO)~\cite{Irastorza:2011gs,Irastorza:1567109}. The IAXO proposal relies on a purposely built large toroidal magnet~\cite{Shilon:2012te} that surpass the CAST magnet figure of merit by substantially increasing the cross-sectional area of the magnet (that goes from $\sim$ 30 cm$^2$ up to $\sim$2~m$^2$). In order to increase correspondingly the sensitivity of the experiment, all the area must be coupled to x-ray focussing systems. Finally, IAXO could benefit from x-ray detectors with background levels as low as 10$^{-7}$--10$^{-8}$~\cskc. These prospects highly motivate the development of x-ray detectors with ultra-low background levels.

\medskip
CAST has already enjoyed a sustained effort in lowering detector backgrounds over the last years. The background level achieved in the latest sunset Micromegas detectors in the 2012 CAST data taking campaign has been of 1.5$\times$10$^{-6}$\cskc~\cite{ThPapaevangelou:SPSC2012}, about 4--5 orders of magnitude below the raw level of a naked detector, and a factor 100 of improvement with respect the first generation of detectors used at the beginning of the experiment. The technology of \textit{microbulk} Micromegas has proven a successful one in implementing a number of developments along three main lines: 1) a light and clean (from the radioactivity point of view) material budget; 2) development of powerful offline discrimination algorithms on the registered topology of events in a gaseous conversion volume, thanks to the use of highly granular readouts; and 3) the design of dedicated passive and active shielding, particularly matched to the gas detection medium.

\medskip
These strategies are conceptually similar to the low background techniques heavily developed in the context of underground experiments (WIMP dark matter and double beta decay) and indeed much of this know-how can be used for our goal (e.g. regarding shielding design). However, some important differences should be stressed, and taken into account carefully when extrapolating existing know-how and techniques to our particular situation. The first one is that our experimentation occurs on surface. The conventional role of the shielding in underground experiments does not equally apply here, and one has to expect cosmic ray secondaries in the shielding material. The second one is the low efficiency of our detectors to higher energy gamma radiation (an important source of background in typical low background experiments) and their relatively high level of discrimination. Because of this, the well known shielding saturation thickness produced by cosmic rays secondaries that is mentioned in introductory texts of low background techniques may need to be revised in our case. It is surprising, for example, that properly shielded Micromegas achieve background levels that are close to the ones produced by some detector component's radioactivity, something unthinkable in solid detectors operated at surface. Because of this and other reasons, the quest for an ultra-low background x-ray detector at surface is a particular development.
 
\medskip
In the present paper we describe the basis of Micromegas detectors for low energy x-ray detection (section \ref{sec:workingPrinciple}). We will describe their performance at CAST, including the last improvements and the current state of the art (section \ref{sec:sunset2012}). We will describe the status of the efforts to study and improve the current background understanding, by means of detailed simulations (section \ref{sec:sims}) and operation of test benches (section \ref{sec:tests}). We will show that this technology offers realistic prospects to achieve the required levels for IAXO, and we will present our demonstrative plans towards this (section \ref{sec:prospects}). We finish with our conclusions in section \ref{sec:conclusions}. This development is carried out within the more generic framework of the T-REX project~\cite{Irastorza2011_EAS}, funded by a Starting Grant of the European Research Council, to develop novel concepts of TPC readout for rare event searches. We focus here to the application to axion research, but this development has important links with WIMP dark matter detection (e.g. via imaging of nuclear recoil directionality~\cite{Ahlen:2009ev}) or the search for the double beta decay of $^{136}$Xe~\cite{Cebrian:2010nw}.

\section{Detection of x-rays with Micromegas.}\label{sec:workingPrinciple}

\begin{figure}[htb!]
\centering
\includegraphics[height=0.35\textwidth]{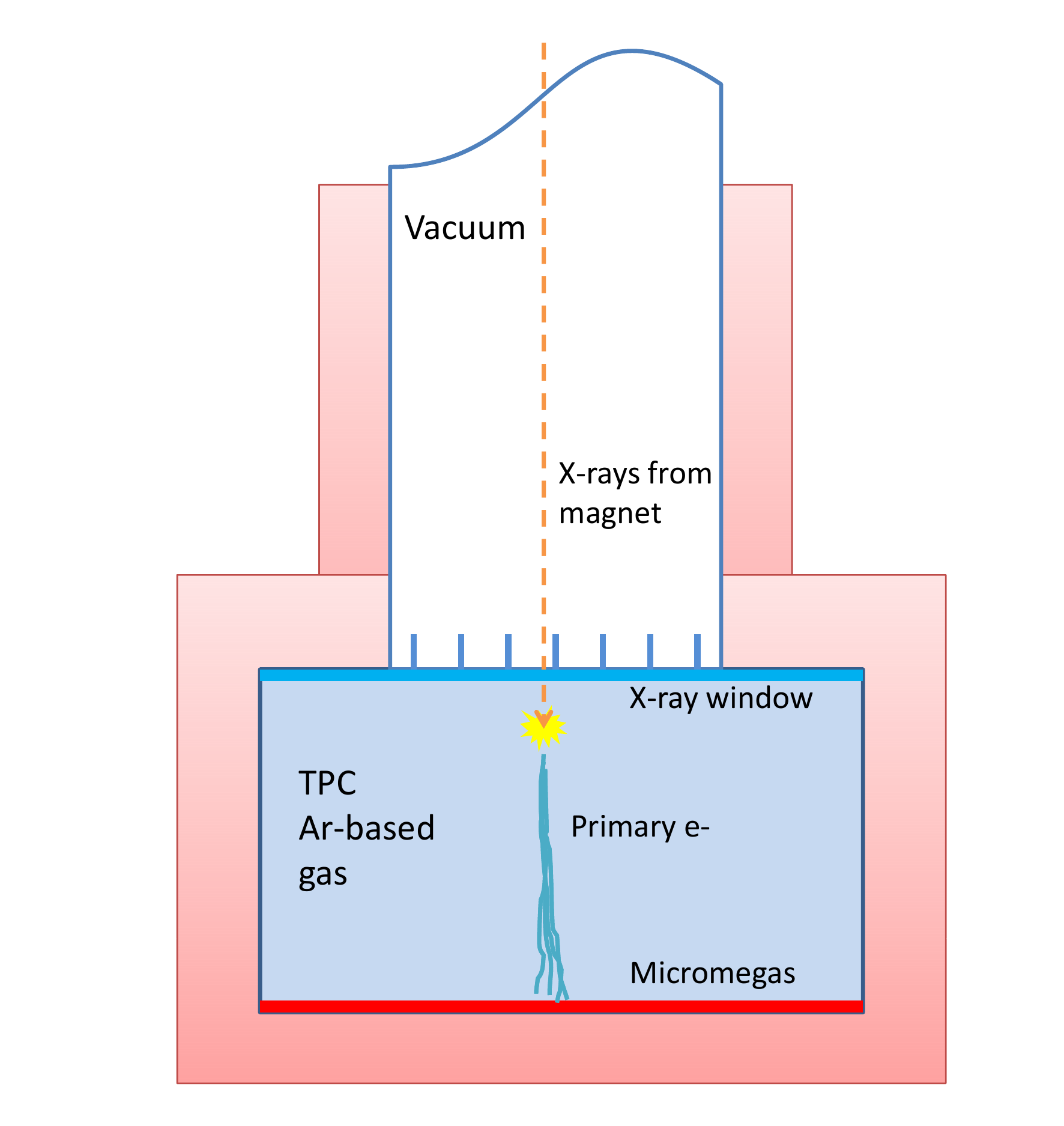}
\includegraphics[height=0.35\textwidth]{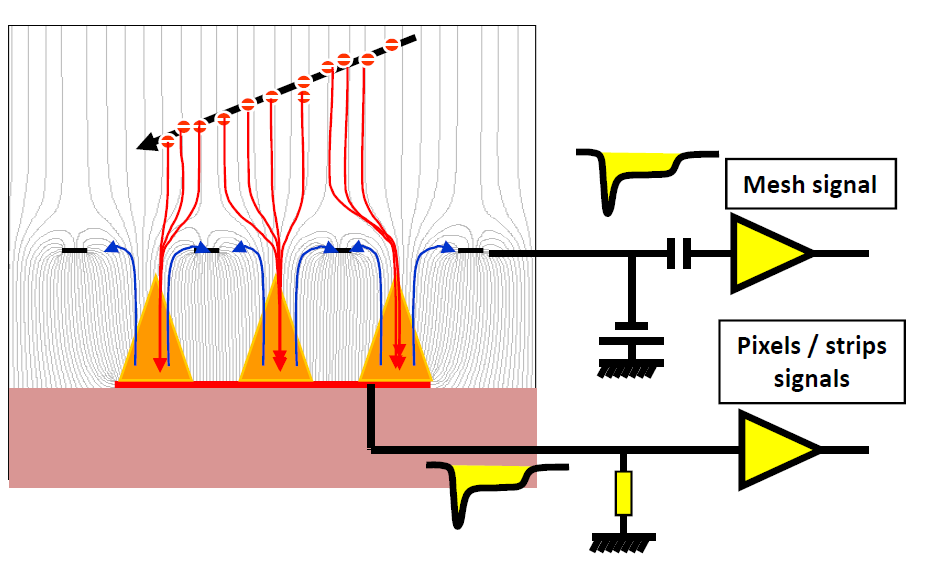}
\caption{TPC-Micromegas usage as x-ray detectors. Left: sketch of the chamber and detection of an x-ray. Right: field configuration and charge amplification in the Micromegas and induced signal in both the mesh and the anode strips.}
\label{fig:workingPrinciple}
\end{figure}

Micromegas are gas amplification structures that can be used as readouts for Time Projection Chambers (TPCs). The working principle of a TPC-Micromegas for x-ray detection is shown in Fig. ~\ref{fig:workingPrinciple}. The TPC cathode is made of an x-ray transparent window. X-rays enter the chamber via this window and interact in the conversion volume of the TPC, usually filled with an Ar-based mixture. The electrons released are drifted towards the Micromegas plane placed at the anode by the electric field present in the conversion volume.


\medskip
The charge amplification occurs in the \emph{amplification gap} of the Micromegas~\cite{Giomataris:1995fq} (see Fig.~\ref{fig:workingPrinciple} right) formed by a micromesh suspended over the anode by means of dielectric pillars. The electric field in the Micromegas gap, about 100 times more intense than the drift field, causes the primary electrons to go through the holes of the micromesh and trigger charge avalanches, including detectable signals in both electrodes. In the microbulk Micromegas~\cite{Andriamonje:2010zz,Iguaz:2012ur}, the readout plane and the amplification structure are developed as a single unity. The structure is manufactured from double-clad kapton foils: the mesh is etched out of one of the copper layers of the foil, while the amplification gap is created by removing part of the kapton by means of chemical and photolitographic techniques. This technique is known to yield the highest precision in the gap homogeneity and, because of that, the best energy resolutions among Micropattern detectors. Typical dimensions are 50$\,\mu$m for the gap and 5$\,\mu$m thickness for the micromesh.

\medskip
The Micromegas anode can easily be patterned with high granularity. Given the typical dimensions of x-ray induced primary ionization clouds, sub-mm granularity is required for efficient signal identification and background rejection. The efficiency of detection of x-rays will be determined by their probability of interaction in the conversion volume, and therefore on its thickness, density and gas composition. The chamber's height is chosen to be a compromise that optimizes the quantum efficiency and minimizes the undesired effects derived from the drift of electrons towards the amplification structure (diffusion and attachment). The possibilities of working at pressure several times higher than atmospheric or with more efficient xenon-based mixtures have been tested to be feasible~\cite{Dafni:2009jb,Cebrian:2012sp}. In addition, the transparency of the chamber windows to x-rays and the signal efficiency of the offline discrimination algorithms applied on the event topology and waveforms will somehow reduce the final detection efficiency. The latter is related with the Micromegas signal to noise ratio (SNR), and the readout pattern.

\medskip
The generic strategies that Micromegas detectors rely on to achieve low background x-ray detection are listed below, while their detailed implementation and the impact on the CAST detectors background will be explained in the following sections.

\begin{itemize}

\item \textbf{Manufacturing technology:} the progressive improvement of the performance of the Micromegas detectors in CAST is closely linked to the development of the Micromegas technology. The CAST experiment has been one of the most demanding test-bench for Micromegas detectors and, in return, CAST has readily benefited from Micromegas detector achievements. Since its introduction in CAST in 2007, the microbulk technology has become technically consolidated, and it has shown to enjoy several advantages over more conventional techniques regarding low background strategies.

\item \textbf{Radiopurity:} intrinsic radioactivity in the detector materials may be a source of background. Microbulk readout planes are made of kapton and copper, two materials of well known radiopurity. Indeed their intrinsic radioactivity have been strongly constrained by a series of dedicated measurements carried out with a high purity Ge detector in the Canfranc Underground Laboratory (LSC) \cite{Cebrian:2010ta}. The geometry of the chamber is relatively simple and its components (chamber's body, x-ray window, screws, gas gaskets, connectors, etc.) have gone through screening campaings and are built up from radiopure materials.


\item {\textbf{Off-line rejection algorithms:} The detailed information obtained by the patterned anode, complemented with the digitized temporal wave-form of the mesh, is the basis to develop advanced algorithms to discriminate signal x-ray events from any other type of events. The power of this discrimination is highly coupled to the quality of the readout so that improvements in readout design or manufacturing yield improvements in discrimination power. The same stands for the front-end electronics and acquisition system.}

\item{ \textbf{Shielding:} Although shielding concepts from underground experimentation can be borrowed, care must be paid to the specifics of our case, e.g., the space and weight constraints of the magnet moving platform, the operation at surface (presence of cosmic rays), the geometry imposed by the magnet (the shielding will always have an opening from which the signal x-rays reach the detector) and the intrinsic sensitivity and rejection capability of the Micromegas detectors.}

\end{itemize}

\section{The Micromegas detectors of CAST.}\label{sec:CASTMM}

In this section we describe the detectors geometry, setup and methods that were used in the CAST microbulk detectors since 2008 to 2012. They led to background levels of about $5-7 \times 10^{-6}$ \ckcs  with a signal efficiency of about 75--90\% at 6 keV ~\cite{Aune:2009zz,mmcast_mpgd09,XRS:XRS1319}. This level can be clearly identified in the CAST Micromegas background history plot, presented in Fig. \ref{fig:histoPlot}. The background energy spectra achieved by these generation of detectors can be seen in Fig. \ref{fig:spectraEvolution}; it is characterized by a copper fluorescence peak (mostly outside the RoI), a small argon fluorescence peak and an important accumulation of counts around the 5--7 keV region, which dominates the level in the RoI.

\begin{figure}[htb!]
\centering
\includegraphics[height=0.37\textwidth]{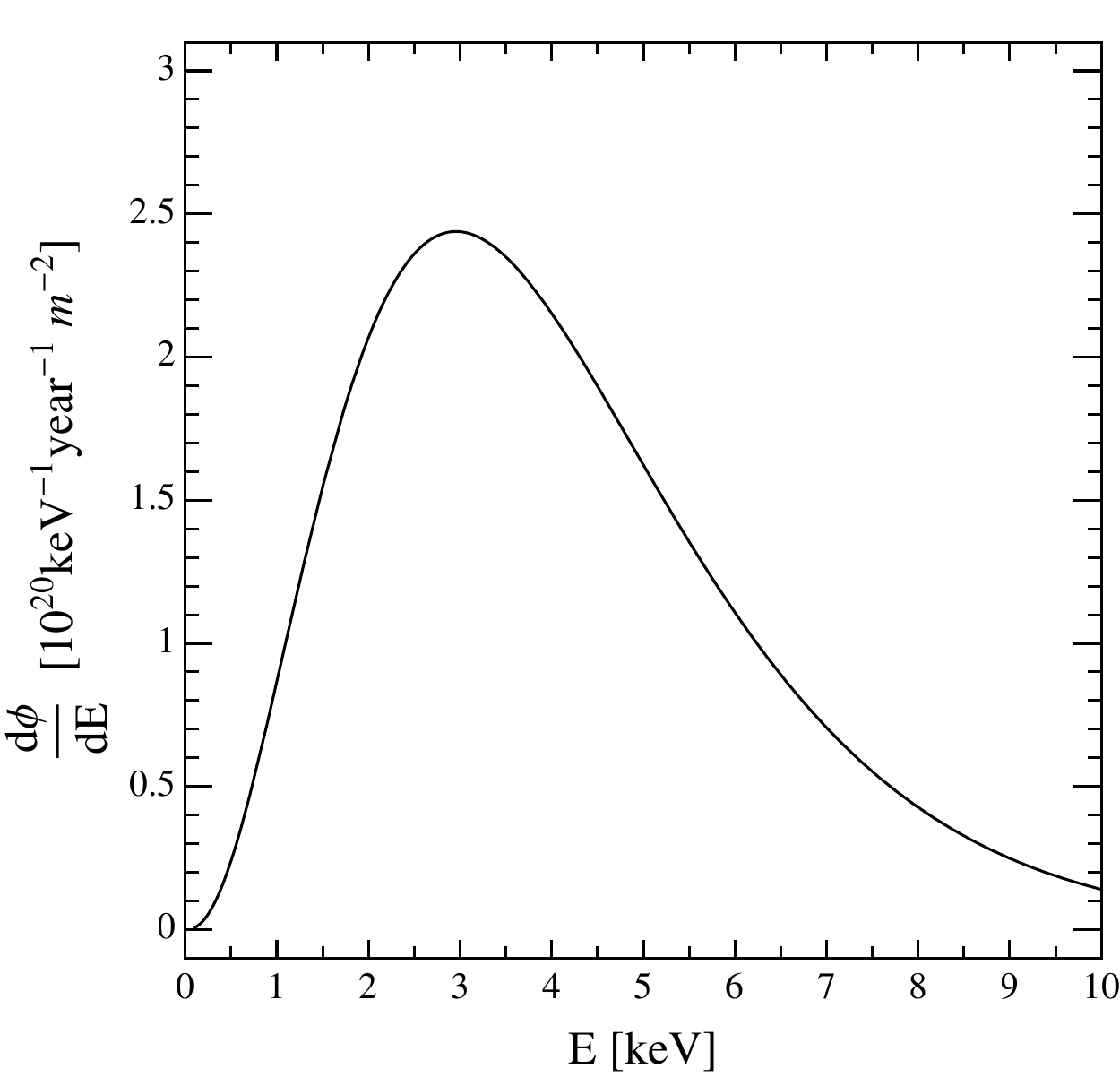}
\includegraphics[height=0.37\textwidth]{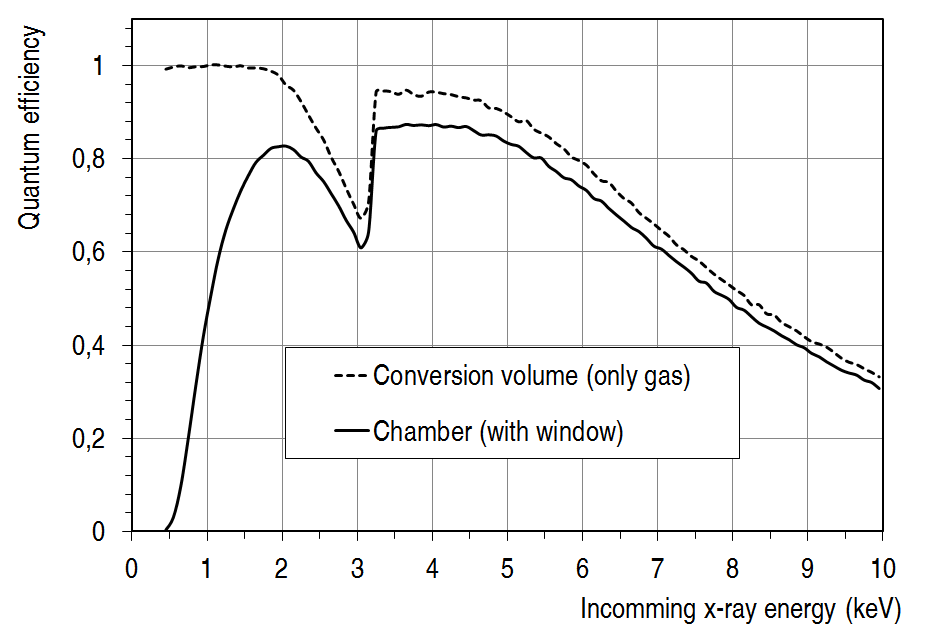}
\caption{Left: expected solar axions flux spectrum at Earth as calculated in ~\cite{Bahcall:2004fg} corresponding to the most generic situation in which only the Primakoff conversion of plasma photons into axions is assumed. Right: quantum efficiency computed from Monte Carlo simulations (using the Geant4 model described in section 4) for only the conversion volume filled with argon with 2.3\% of isobutane at 1.4 bar and with the addition of the chamber's x-ray window with the strong-back grid.}
\label{fig:Qeff_RoI}
\end{figure}

\medskip
The detector's structure has remained basically unchanged since the beginning of the CAST experiment. The conversion volume of the Micromegas chamber has 3 cm height and is filled with argon with 2.3\% of isobutane at 1.4 bar. The x-rays coming from the magnet enter the conversion volume via a gas-tight window made of 5$\mu$m aluminized mylar foil. This foil is also the cathode of the TPC, and it is supported by a metallic squared-pattern strong-back, in order to withstand the pressure difference with respect to the magnet's vacuum system. The effect of foil and strong-back on the detector efficiency is shown on the right of Fig. \ref{fig:Qeff_RoI}.

The expected solar flux of axions extends practically up to 10 keV (see Fig. \ref{fig:Qeff_RoI} left), however the loss of efficiency at low energy and the typical presence of a copper fluorescence peak at 8 keV leads to the definition of the CAST RoI from 2 to 7 keV, which encompasses the 74\% of the total axion flux. Although there is motivation, good prospects and work ongoing to reduce the threshold well below 2 keV, we will not review it in this paper, which is focused on low background in the stated RoI. The Micromegas active area is a 36 cm$^2$ square that comfortably covers the 14.55 cm$^2$ projection of the magnet's bore, which is the analysis fiducial area.

\begin{figure}[ht!]
\centering
\includegraphics[height=0.28\textwidth]{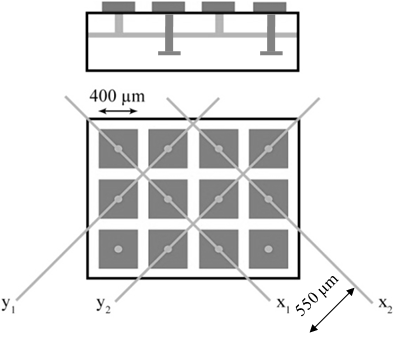}
\includegraphics[height=0.28\textwidth]{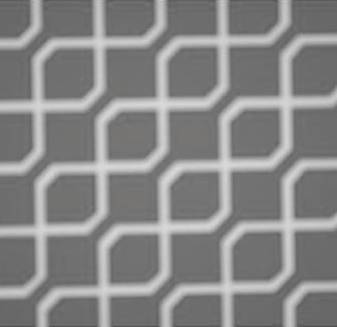}
\includegraphics[height=0.28\textwidth]{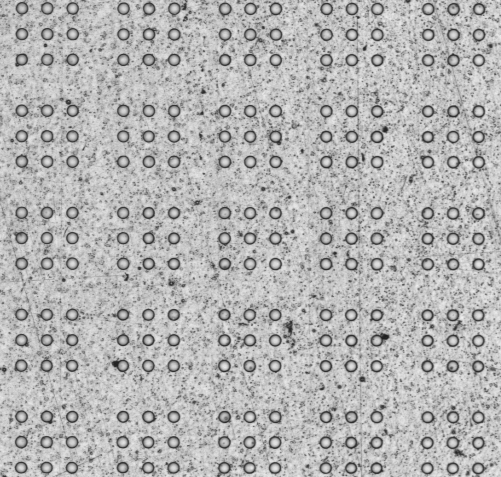}
\caption{CAST microbulk's readout design. Left: anode layout. Right: mesh pattern, the groups of holes are correlated with anode's pads.}
\label{fig:lastReadouts}
\end{figure}

The pattern of the Micromegas anode follows the schema represented on the left of Fig. \ref{fig:lastReadouts}, with an array of highly granular pixels interconnected in $x$ and $y$ directions. An implementation of this schema can be observed in the middle of Fig. \ref{fig:lastReadouts}, where half of the pixels are connected by short and narrow strips in the anode plane (thus reducing the material budget), and the other half are connected in an underlying copper plane. The holes of the micromesh are arranged in groups that follow the pixels underneath, as shown on the right of the Fig. \ref{fig:lastReadouts}.

 
\medskip
The signals from the strips are amplified and integrated with a Front-end Gassiplex card\cite{Santiard:1994ps} controlled by a CAEN sequencer (V 551B) with two CRAM (V550) modules in a VME crate\cite{StripsDAQCASTclasic}. With such electronics a 10 bit value is available for each one of the $2\times106$ strips. The mesh pulse, used as trigger signal to acquire the strips, is also sampled with a VME digitizing board, the MATACQ (MATrix for ACQuisition)\cite{MATACQ} 12 bits dynamic range, at 1 GHz sampling frequency and an acquisition window of 2.5 $\mu$s per event.

\subsection{Characterization and performance of CAST microbulk Micromegas.}\label{sec:Performance}

\begin{figure}[b]
\centering
\includegraphics[height=0.32\textwidth]{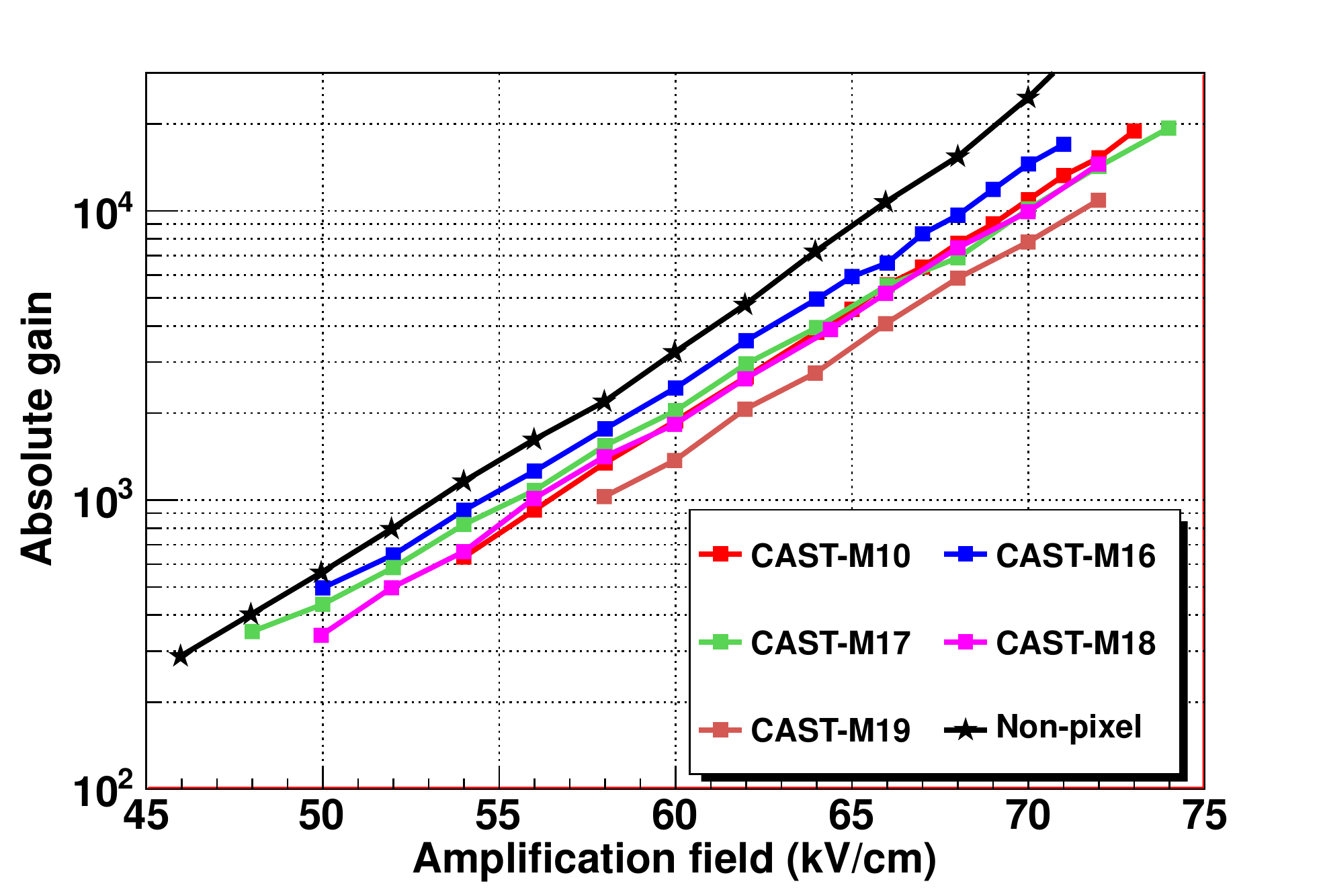}
\includegraphics[height=0.32\textwidth]{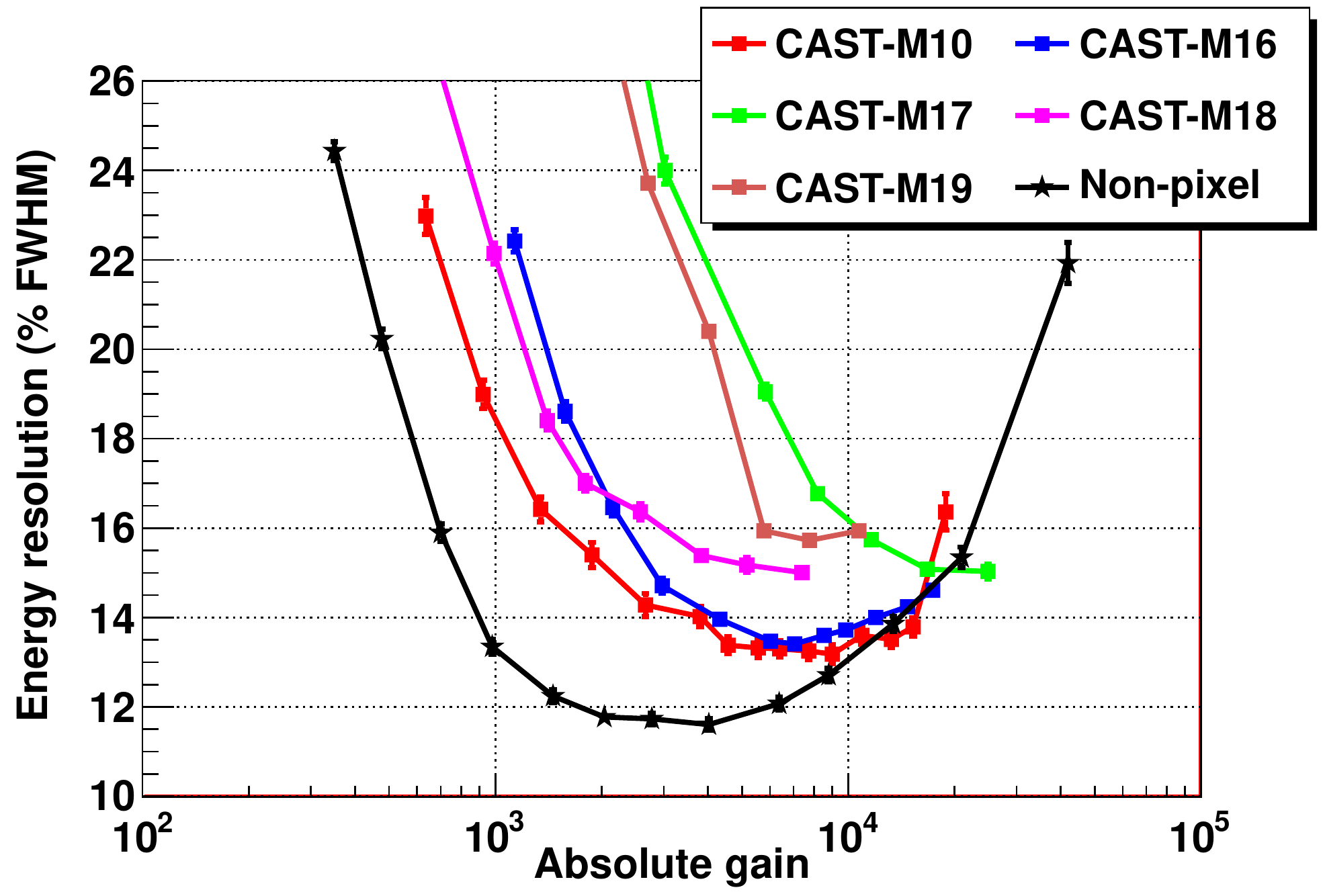}
\caption{Dependence of the gain with the amplification field (left) and of the energy resolution at 5.9 keV with the gain (right) obtained from mesh signals for different CAST detectors in Ar+5\%iC$_4$H$_{10}$ at atmospheric pressure. As a comparison, the curves generated by a non-pixelized microbulk detector \cite{Iguaz:2012ur} in the same conditions (black stars) have been included.}
\label{fig:GainEResCAST}
\end{figure}

The Fig. \ref{fig:GainEResCAST} summarizes systematic characterizations of CAST microbulk detectors carried out in the same test conditions: using argon with 5\%iC$_4$H$_{10}$ at atmospheric pressure and shaping the mesh signal with 1$\,\mu$s shaping constant to generate the energy spectrum. These are standard test conditions to perform comparisons of performance between Micromegas, however they differ slightly from the normal CAST operation conditions mentioned before.

\medskip
All the CAST detectors reach gains as high as 10$^4$ from the mesh signal and energy resolution values between 13--16\% FWHM are routinely obtained at 5.9 keV irradiating the whole Micromegas area. Those values are not far from the 11\% FWHM reached by smaller non-pixelated microbulk prototypes (3.5 cm diameter)~\cite{Iguaz:2012ur}. Better values down to 12.5 \% FWHM are systematically obtained when the signal from the strips is used, what could be an indication that the previous results are limited by noise conditions rather than being intrinsic limitations of the readouts (e.g. gap inhomogeneities over the surface).

\begin{figure}[htb!]
\centering
\includegraphics[width=0.85\textwidth]{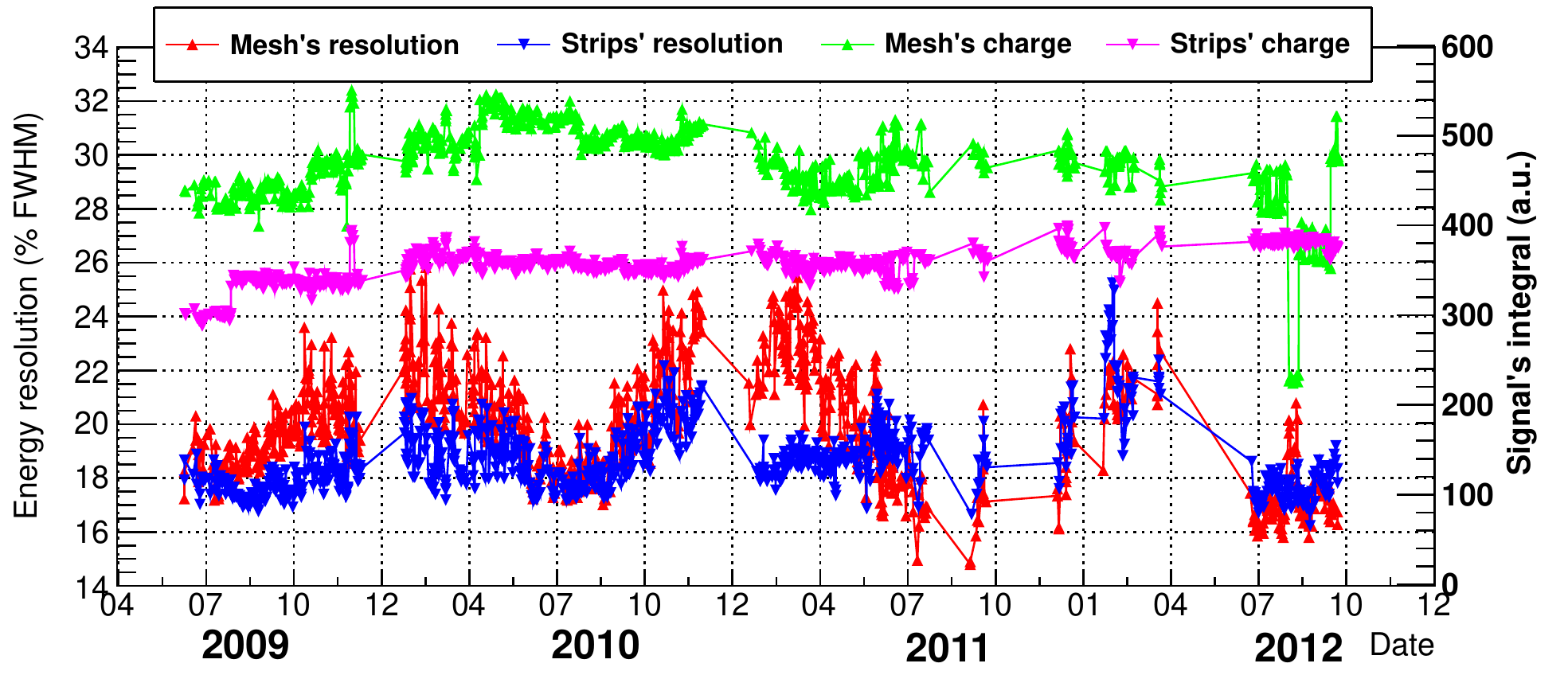}
\caption{Mesh and strips gain and energy resolution of the same sunrise microbulk detector during more than 3 years in operation in CAST. Gaps corresponds to the CAST shut-down periods. Sharp changes are related with the variation of the detector operation parameters, mainly the mesh voltage. Strips and mesh gain are not necessarily correlated since the electronics configuration of the latter can be changed too.}
\label{fig:EvolCASTSunrise}
\end{figure}

\medskip
Long term stability and reliability is an imortant requirement for every detection technology to be used in rare event searches. Microbulk technology has demonstrated to fulfill these requirements. The same CAST microbulk detector has been operated during more than three years in one of the sunrise docking points. Its main performance parameters, both strips and mesh gain and energy resolution, throughout this period are plotted in Fig. \ref{fig:EvolCASTSunrise}. The energy resolution of the detector in operation in CAST is worse than during the characterization tests (Fig. \ref{fig:GainEResCAST}, right) due, in part, to the variation of the iC$_4$H$_{10}$ proportion and pressure of the gas, also to the shortening of the amplifier shaping time in order to improve the temporal information, and to the fact that the detector is operated in a lower-than-optimal gain, to increase the safety margin against potentially damaging discharges.

\subsection{Event selection with CAST Micromegas.}\label{sec:Discrimination}

The basic idea on which the CAST analysis relies is that x-ray events inside the RoI are seen as small and symmetric clouds by the  Data Acquisition System (DAQ) due to the fact that the ionization range of the  photo-electron is short compared with the typical electron diffusion lengths. In contrast, other ionizing particles, like muons or electrons of higher energy producing energy depositions inside the RoI, should be identified because of their different ionization topology.

\medskip
In order to apply the discrimination criteria, the signals recorded with the DAQ are reduced to a series of parameters. The pulse shape analysis of the digitized mesh signal offers information on the extension and structure of the event along the $z$ axis (drift) direction, while the $x$ and $y$ projections of the event topology is given by the strips signals. X-rays tipically generate a single compact group (cluster) of fired strips. In principle, events with only one cluster per axis are accepted.  From these basic considerations, a series of analysis observables are derived. They are basically of three types:


\begin{itemize}
\item Pulse shape observables: like risetime, risetime/width and amplitude/integral. 

\item Event size observables: like the maximum cluster size and the balance between the cluster size on X and Y directions.

\item Observables that represent balances between mesh and strips information: like ratio between the pulse integral of the mesh and the clusters total charge recorded by the strips.  
\end{itemize}

\begin{figure}[htb!]
\centering
\includegraphics[height=0.28\textwidth]{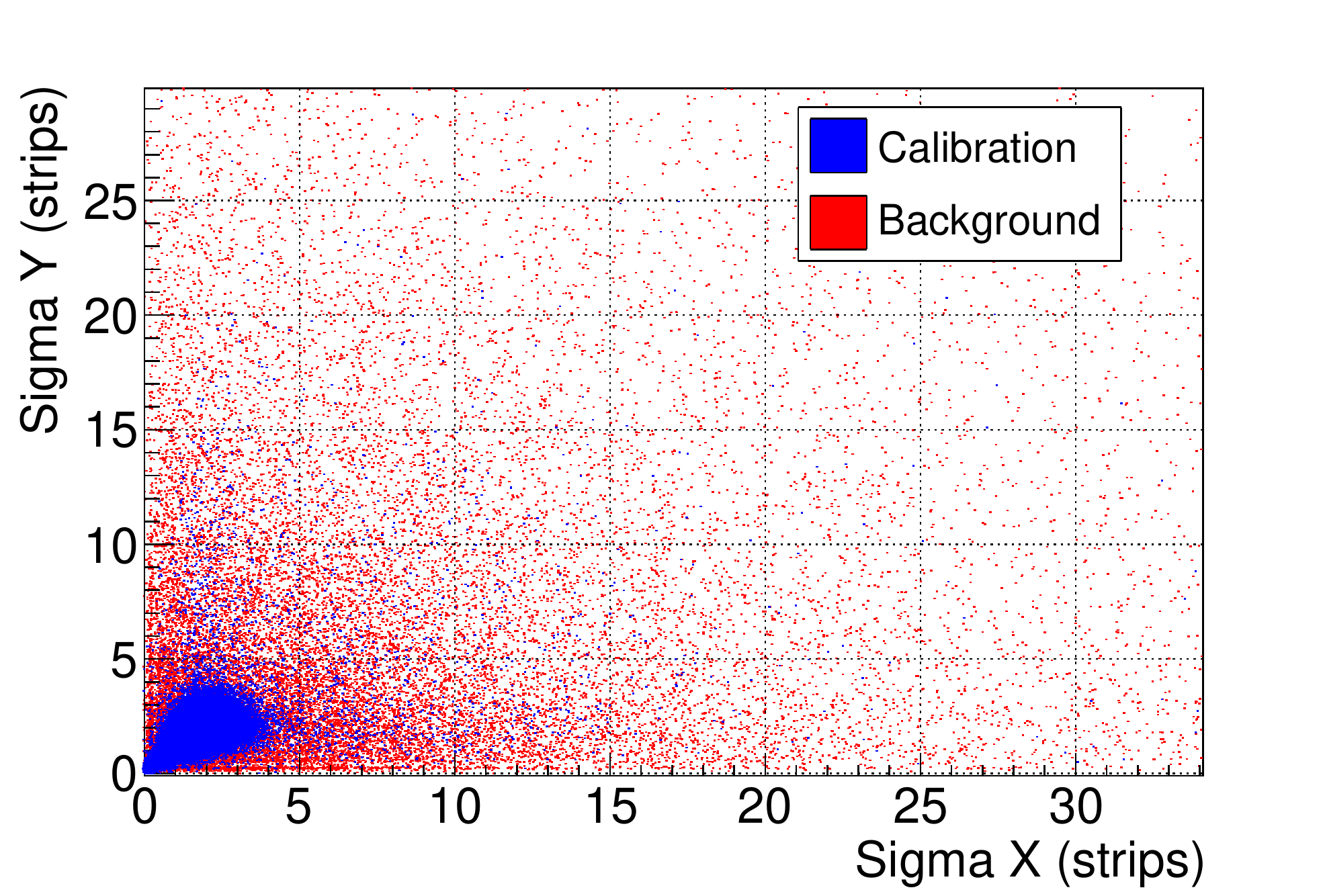}
\includegraphics[height=0.28\textwidth]{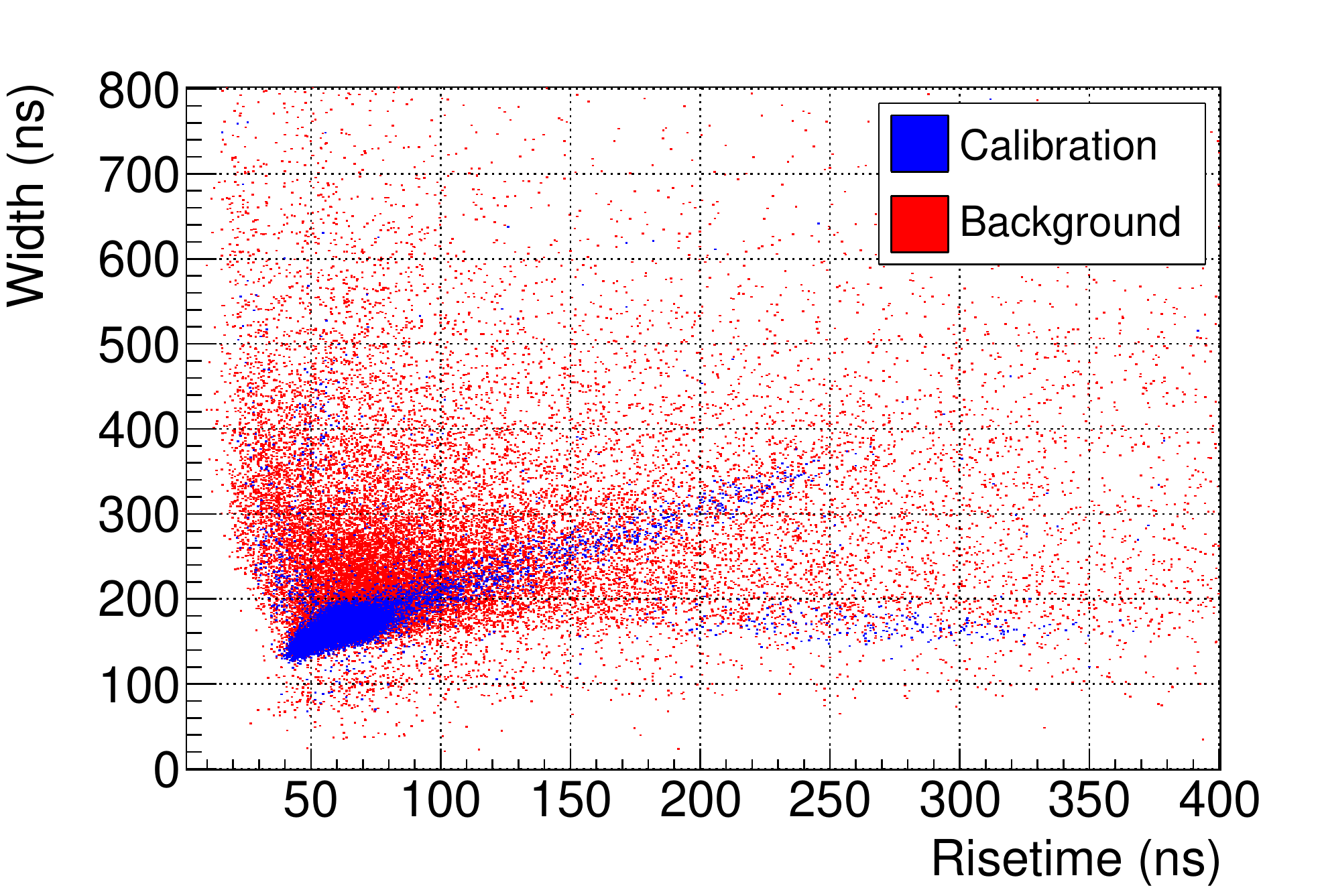}
\caption{Event distributions from calibration (blue) and background (red) events compared for two couples of observables: on the left, pulse shape observables, width vs risetime. In the middle, strips-based observables, cluster's size $x$ vs size $y$, computed as r.m.s.}
\label{fig:discrimination}
\end{figure}

\begin{figure}[htb!]
\centering
\includegraphics[height=0.35\textwidth]{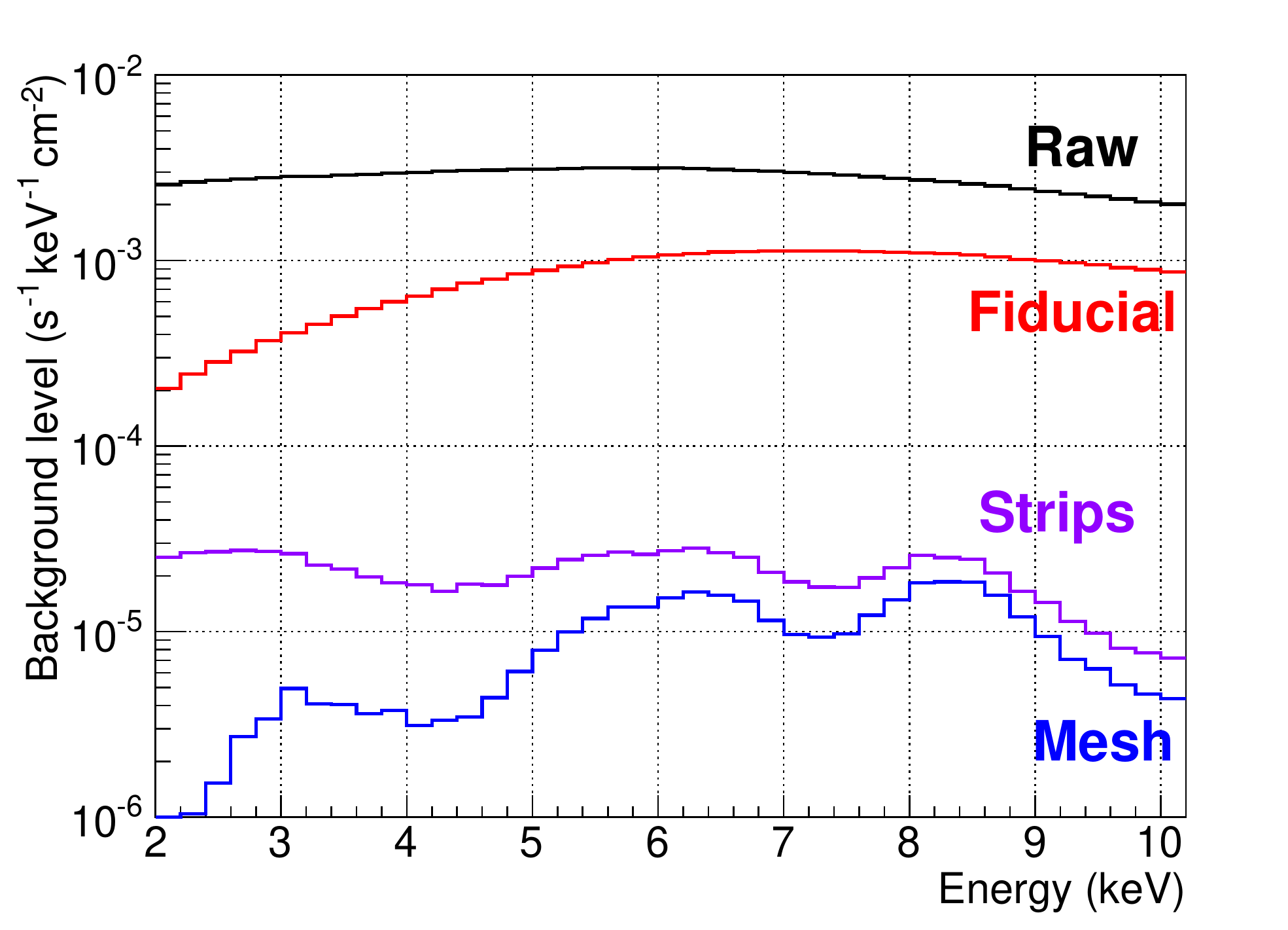}
\caption{Background suppression of the raw background spectrum (black) after surface fiduialization (red), and applying the selection criteria related with the strips (purple) and mesh (blue) observables. }
\label{fig:bkgsupression}
\end{figure}

\medskip
The discrimination criteria, and the signal efficiency, are defined from the populations of 6 and 3 keV (argon scape peak) x-rays from a calibration with a $^{55}$Fe source. The detectors are daily calibrated to avoid any possible systematic effect related with performance degradation or variation of settings or environmental conditions in time. In the Fig. \ref{fig:discrimination}, the observables' distribution of calibration and background events are confronted. A detailed description of the discrimination algorithms is out of the scope of this work. We will just mention that, typically, a set from 6 to 10 observables are involved in the discrimination criteria, which can be applied sequentially~\cite{Iguaz:2011xj} or using multivariate analysis techniques~\cite{mmcast_mpgd09}. In Fig. \ref{fig:bkgsupression}, we illustrate the effect of sequentially applying fiducialization, strips and mesh discrimination criteria to the raw background events.

\subsection{Shielding description: sunrise setup 2007--2012.} \label{sec:setupdescription}
The shielding approach described here was applied to the CAST Micromegas detectors in 2007, and lasted until the progresses reported in this work motivated a new upgrade in 2012.

The shielding was a compromise between the application of the different shielding strategies against most of the potential external background sources (whose contribution was unknown) and the space constrains of the magnet platform. The main shielding material is a 2.5 cm thick layer of archeological lead. Inside the lead there is a 5 mm thick layer of pure copper to stop low energy radiation generated in the lead. Outside the lead, a neutron shielding is placed, composed of polyethylene bricks (variable thickness, up to 10 cm) followed by a 1 mm layer of cadmium to absorb thermalized neutrons. The innermost three layers (cadmium, lead and copper) are enclosed in a cylindrical plexiglass box which, moreover, is flushed with nitrogen to avoid radon concentrations close to the detector. The shielding has an aperture used for the signal extraction (signal outlet).


\medskip
In order to reduce the background level, the CAST Micromegas team undertook a study on the nature of the background of CAST Micromegas. In particular, a new design of the detector's shielding was conceived. The associated R\&D program includes both Monte Carlo simulations and test bench activities, whose present status will be reported in the next sections.

\section{Monte Carlo simulation of the CAST Micromegas setup}\label{sec:sims}

The CAST Micromegas setup was modeled in Geant4 \cite{Agostinelli:2002hh} in order to estimate the different background contributions. The most important study was the effect of the external gamma flux from the environmental radioactivity like e.g. that of the concrete walls of the experimental hall. The simulation was performed for many individual initial photon energies and with directions impinging isotropically the simulated geometry. The results are then weighted using experimental data taken in-situ with a NaI detector, to represent the real external photon flux at CAST. This procedure has a large uncertainty for energies below 200 keV.

\begin{figure}[htb!]
\centering
\includegraphics[width=0.95\textwidth]{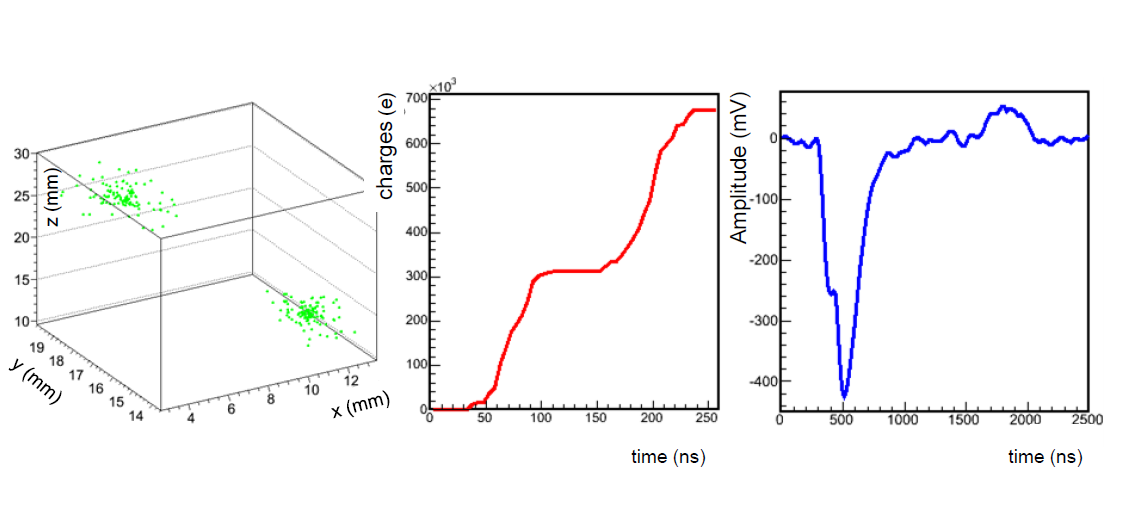}
\caption{Some of the different stages of the Monte Carlo chain. First: electron clouds after primary charge generation for two 3 keV photons from a $^{55}$Fe calibration event, including the diffusion effects during the drift through the CAST chamber. Second: charge collection and amplification in the Micromegas. Third: signal generated in the shaper-amplifier, including electronic noise.}
\label{fig:simChain}
\end{figure}

\medskip
The data simulated was treated with the RESTsoft package, a generic purpose framework to deal with data from TPCs in rare event applications, developed at the University of Zaragoza \cite{alfredotesis}. The simulation chain starts with the generation of the primary electrons from Geant4 energy depositions, which depend on the experimental W values and the Fano factor of the gas mixture considered. In a second step, each electron is drifted to the Micromegas plane, using the parameters extracted from Magboltz\cite{Biagi:1999nwa} (drift velocity, gas diffusion and attachment probability). The probability of passing through the mesh and the avalanche size are then simulated using the experimental characterization of CAST micromegas detectors. Finally, a simple model for the shaper amplifier response to fast signal is used to generate the mesh pulse and the strips signals. In Fig. \ref{fig:simChain}, we show the sequential transformation of a typical energy deposition in the detector, from primary electron generation to the mesh pulse formation.  The model was adjusted by comparing real and simulated calibration data. A direct comparison of a simulated and experimental calibration has been published in \cite{XRS:XRS1319} and the whole simulation chain is described in detail in \cite{alfredotesis}. Afterwards, simulated data is transformed into the actual CAST data format, so that the same analysis and discrimination algorithms than for real data are applied.

\begin{figure}[t]
\centering
\includegraphics[height=0.3\textwidth]{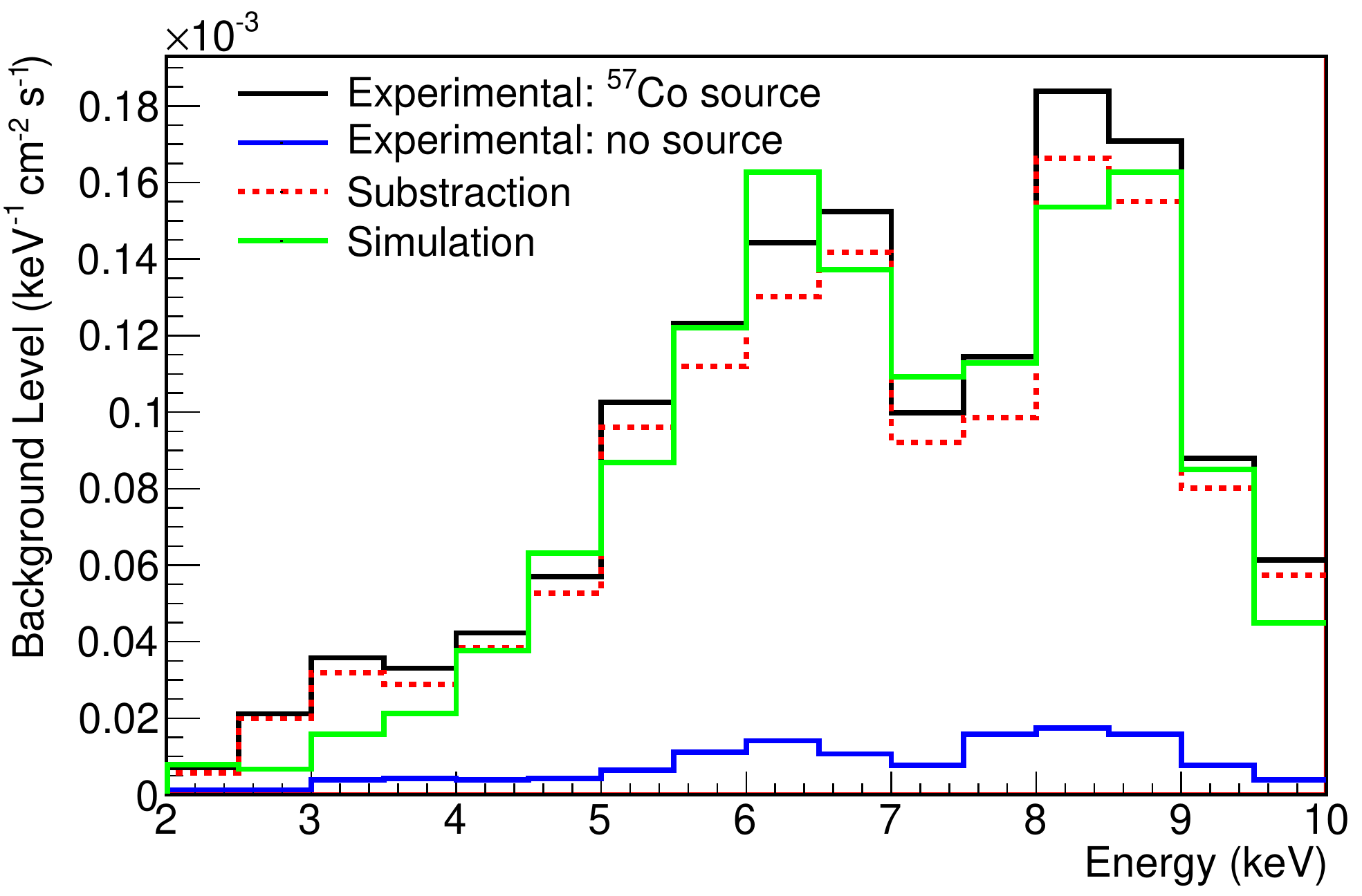}
\includegraphics[height=0.3\textwidth]{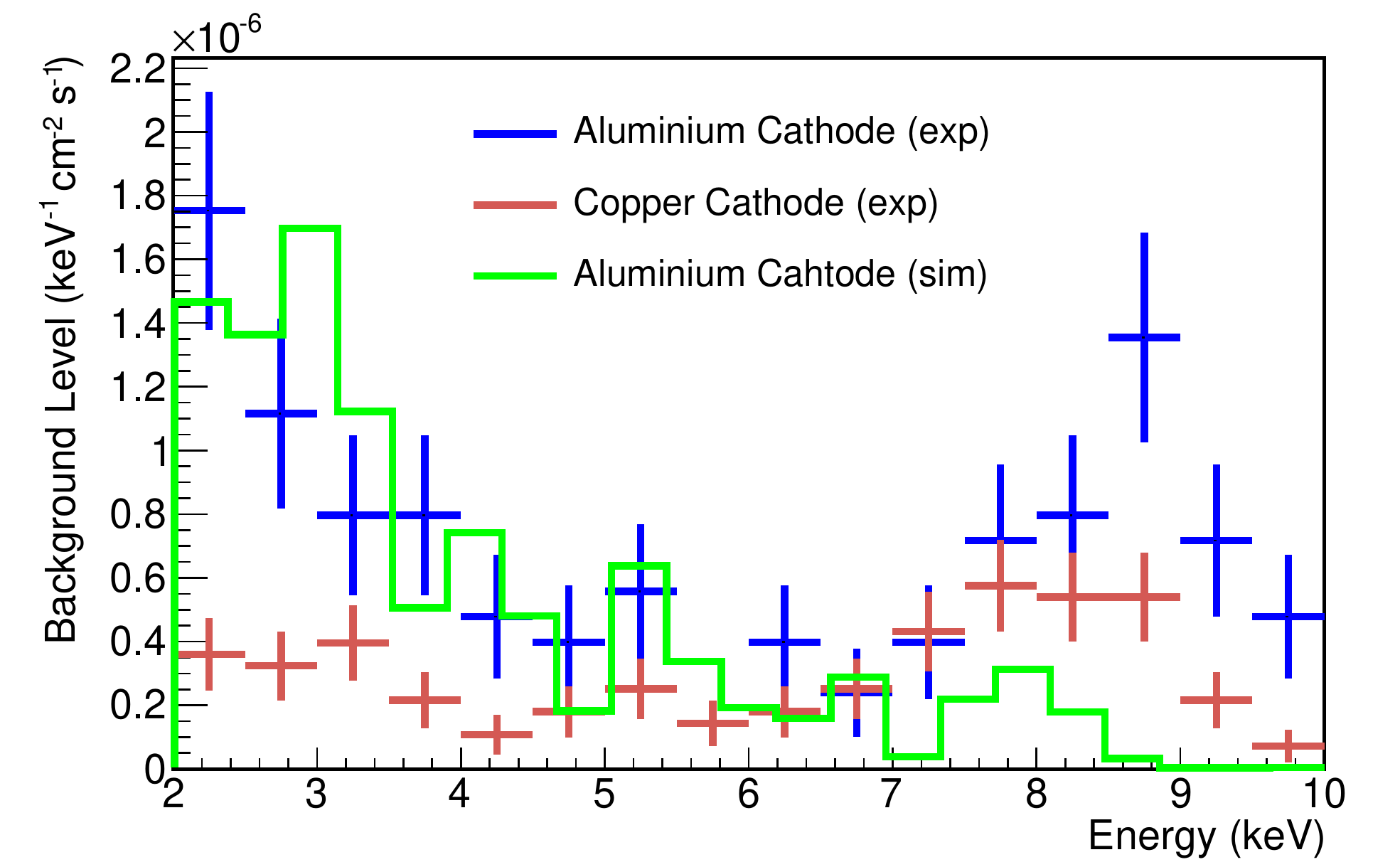}
\caption{Left: Simulation vs experiment for one of the positions of the $^{57}$Co source near the sunrise detector. The source contribution is computed from the subtraction of the nominal background, the difference should be directly compared with the simulated result. Right: the plot shows how the simulation of the background induced by an aluminium cathode (whose activity was measured by germanium spectroscopy) can explain the difference between the background spectra obtained with that cathode and a much more radiopure copper one in the LSC using a 20 cm thick shielding.}
\label{fig:Comparisons}
\end{figure}

\medskip
Simulation and experimental results have been confronted by means of different tests. Several measurements were carried out with a $^{57}$Co source placed in different positions of the Sunrise setup. The tests showed that the 122 and 136 keV gamma lines created a relatively large accumulation of events in the 5--7 keV region, even if no direct path to enter the detector was allowed. The results reasonably agree with the Monte Carlo simulation (see Fig. \ref{fig:Comparisons}, left). This agreement represents an important cross-check fo the overall simulation chain. The same simulation procedure has been used to reproduce the background spectrum induced by th intrinsic radioactivity of an aluminium cathode that was used in a specific test bench (see next section). The radioactivity of the piece was determined in an independent measurement with a high purity Ge detector at the LSC. The agreement between experimental and simulated data is shown in Fig. \ref{fig:Comparisons}. 
This procedure is also used to determine the potential contribution of the remaining detector component's intrinsic radioactivity (e.g. Fig. \ref{fig:Comparisons}, right). From the radiopurity measurements of microbulk readout planes of \cite{Cebrian:2010ta} $\leq10^{-8}$\ckcs is obtained, and from the whole detector structure $\leq3\times10^{-8}$\ckcs, minding the fact that most of the accounted contributions are estimated from upper limits to their radioactivity.


\begin{figure}[htb!]
\centering
\includegraphics[height=0.35\textwidth]{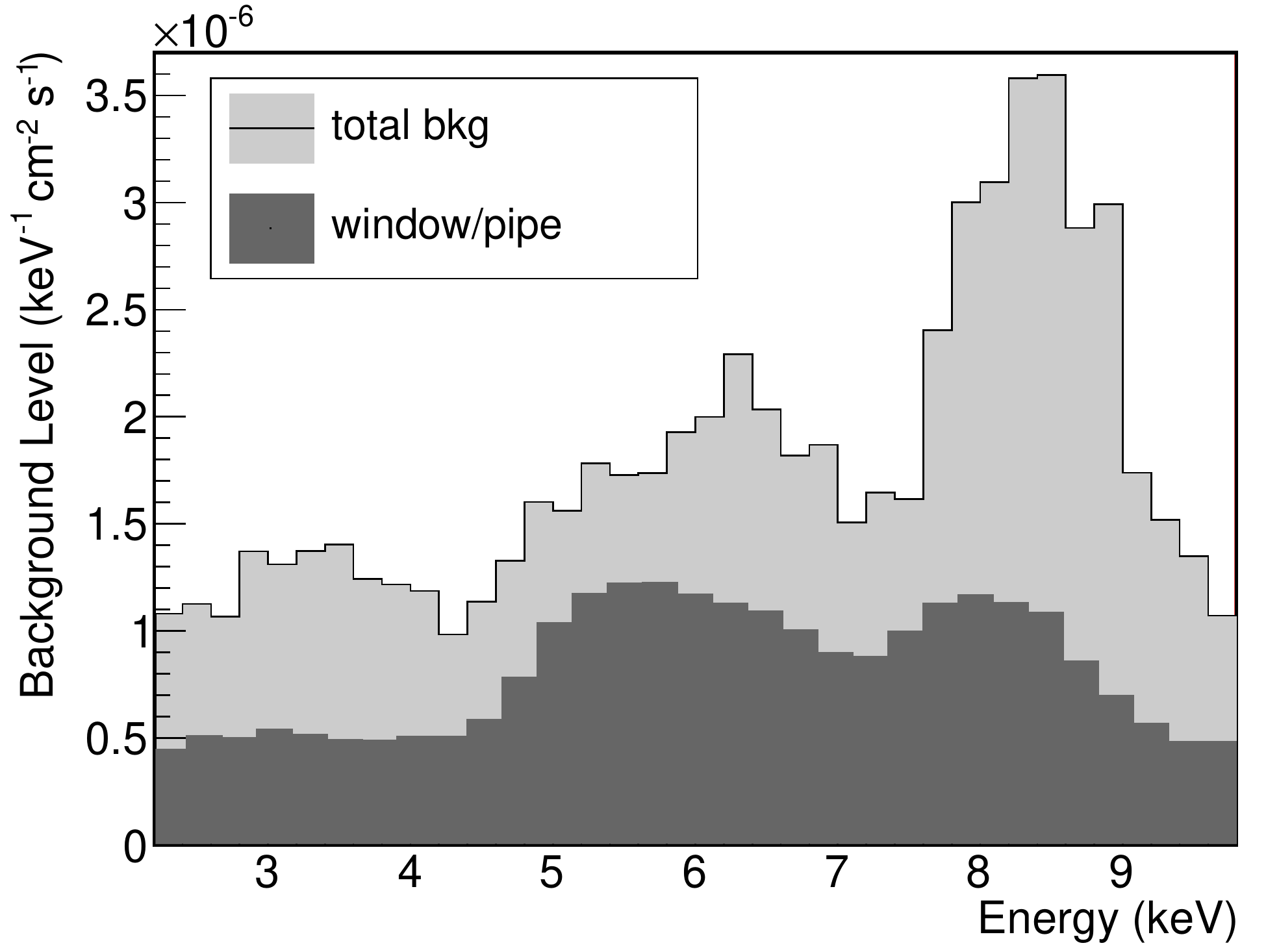}
\includegraphics[height=0.35\textwidth]{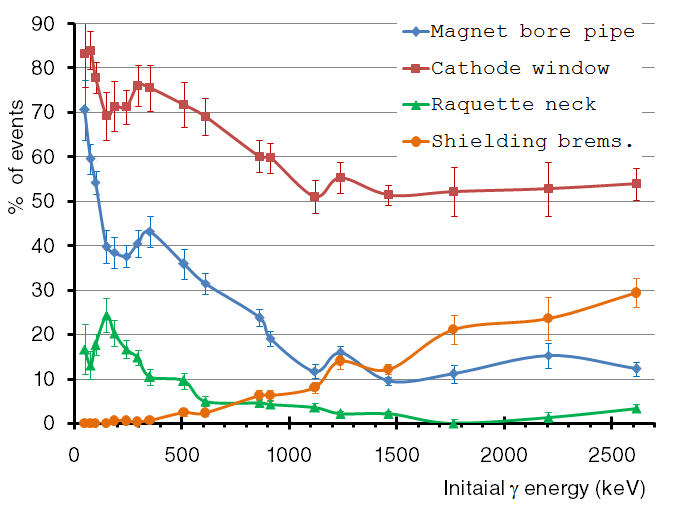}
\caption{Left: final background spectrum obtained from simulation of the CAST area environmental gamma flux. The contribution from intrusion via the magnet pipe and detector window is superposed; it is responsible of 52\% of the total background in RoI and clearly produces the 5--7 keV peak. Right: different mechanisms of production of background counts as a function of the energy of the original gamma, particularly the intrusion through the shielding and detector outlets.}
\label{fig:simResults}
\end{figure}

\medskip
The background level obtained in the RoI after application of the offline cuts to the population of simulated external gamma events properly normalized with the NaI data amounts to $1.5 \times 10^{-6}$\ckcs, about one third experimental level obtained by the CAST sunrise detector, and the resulting energy spectrum is shown in Fig. \ref{fig:simResults} (left). Even if the total experimental background is not reproduced, some qualitative conclusions can be extracted from the dependence of background origin with the initial energy (see in Fig. \ref{fig:simResults}, right). This plot shows that for low energy gammas, most of the events in the RoI were generated by gammas entering through the shielding outlets (magnet's bore, x-ray widnow, signals outlet) instead of penetrating through the shielding or via bremsstrahlung showers. Indeed, more than 50\% of the total were generated by primary or secondary gammas passing through the x-ray window; and, within them, those from scattering in the stainless steel pipe between the detector and the magnet are dominant. These pipe-window events accumulate in the 5--7 keV region due to the fluorescence lines of steel (Cr, Fe and Ni).

\section{Test bench activities}\label{sec:tests}

\begin{figure}[htb!]
\centering
\includegraphics[height=0.35\textwidth]{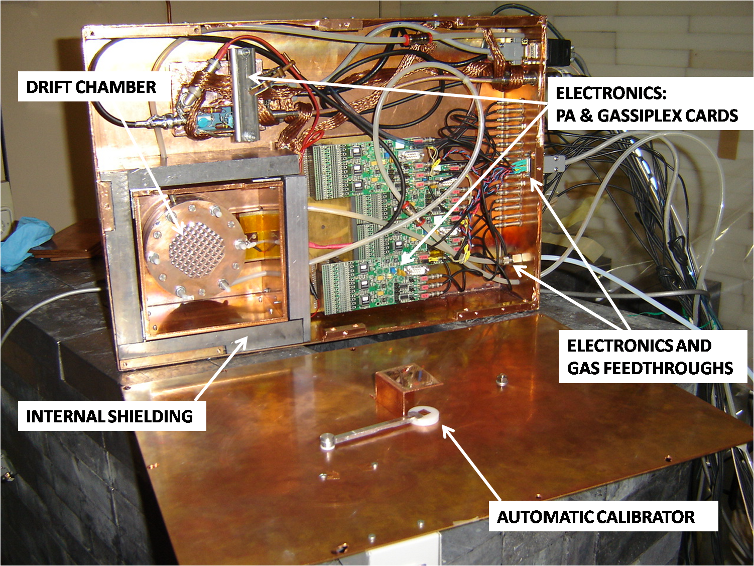}
\includegraphics[height=0.35\textwidth]{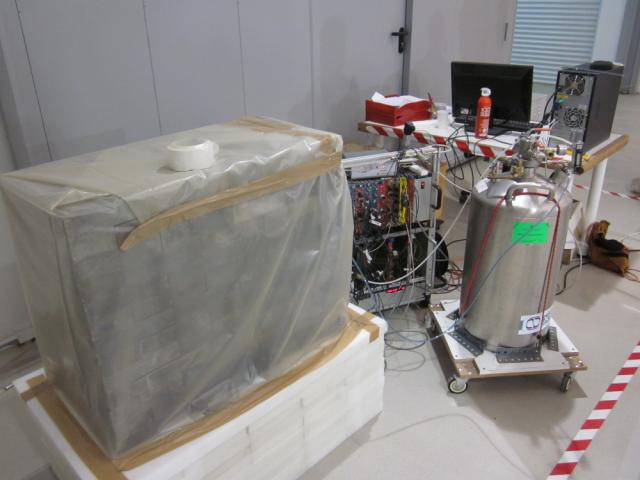}
\caption{Left: CAST sunrise-like setup for tests. Right: The tests setup at the LSC. The basic setup (left) is covered by an external lead layer (10 cm thickness in the photo). To avoid radon concentration nitrogen is flushed inside the Faraday cage and the setup enclosed inside a plastic box. The nitrogen dewar and DAQ electronics are visible.}
\label{fig:set-up}
\end{figure}

The basic CAST micromegas setup described in section \ref{sec:setupdescription} (see Fig. \ref{fig:set-up}, left) has been replicated and installed at surface in the laboratories of University of Zaragoza and underground at the Canfranc Underground Laboratory (LSC). The setup at LSC enjoys very stable environmental conditions and a cosmic muon flux a factor of $10^{4}$\cite{Luzon:2006sh} lower than at surface level. With these setups, an intense program of tests with different shielding configurations has been carried out to evaluate the relative importance of different contributions to the background. Data taken with different amounts of lead, polyethylene or nitrogen flux gives insight on the external gamma, neutron or radon contributions respectively. The contribution of muons can be estimated from the comparison between surface and underground data. Finally, the background attributed to the intrinsic radioactivity of this setup was deduced, once all external components are brought to negligible levels.

A summary table of the most significant background levels achieved underground can be found in table ~\ref{tab:BkgSummary}.

\begin{table}[htb!]
\label{tab:BkgSummary}
\centering
$$
\begin{array}{lcc|cc}
  {\mbox{Detector}}&{\mbox{\hspace{2mm} Ext. shield. thick.}}&{\hspace{2mm} \mbox{ Run time (days)}}&{\mbox{ \hspace{2mm} Bkg. ([2-7] keV)}}&{\hspace{2mm} \mbox{ Bkg. ([2-9] keV)}}\\ 
  \hline
  {\mbox{M10}^\ast}&{\mbox{20 cm}}&{35.0}&{7.1 \pm 1.2}&{7.2 \pm 1.0}\\
  {\mbox{M10}}&{\mbox{20 cm}}&{86.9}&{1.9 \pm 0.3}&{2.1 \pm 0.3}\\
  {\mbox{M17}}&{\mbox{0 cm}}&{6.0}&{67  \pm 8}&{100 \pm 9}\\
  {\mbox{M17}}&{\mbox{5 cm}}&{11.6}&{3.6 \pm 1.4}&{4.8 \pm 1.7}\\
  {\mbox{M17}}&{\mbox{10 cm}}&{43.1}&{2.1 \pm 0.5}&{2.8 \pm 0.5}\\
  {\mbox{M17}^{\ast\ast}}&{\mbox{10 cm}}&{31.6}&{1.1 \pm 0.4}&{1.5 \pm 0.4}\\
  \hline 

\end{array}
$$
\caption{Summary of underground measurements. The thickness corresponds to extra lead coverage, i.e., 0 cm means just like the CAST sunrise setup. M10 was operated in the LSC lab. In particular it is shown the progressive upgrade of the M17 detector in the LSC whose corresponding spectra can be found in Fig. 17. Final background levels expressed in units of $10^{-7}$ \cskc and statistical errors are given as 2$\sigma$. ($^\ast$) measured with Al cathode, rest of measurements using a Cu one. ($^{\ast\ast}$) All remaining steel and brass components replaced with copper and Teflon.} 
\end{table}

\begin{figure}[htb!]
\centering
\includegraphics[width=0.8\textwidth]{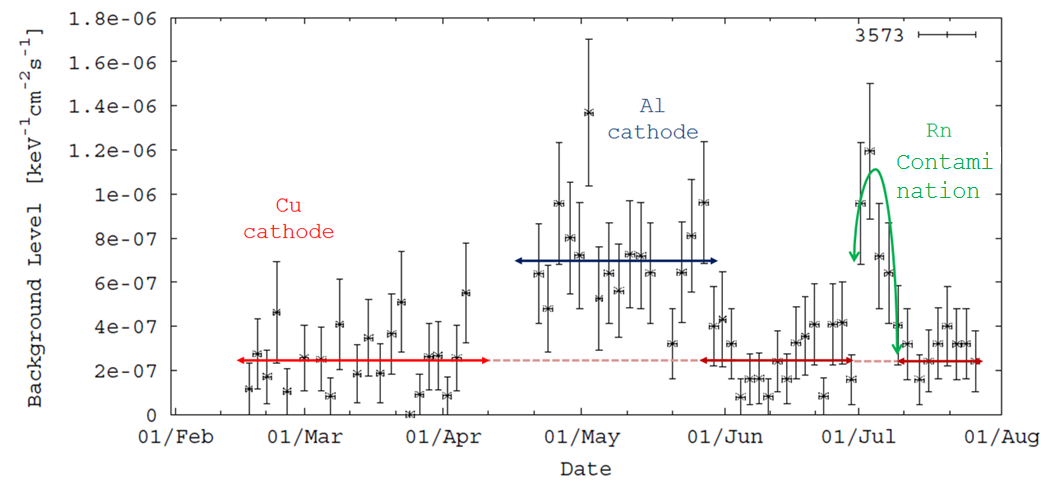}
\caption{Ultra-low background period (2011) twice interrupted because of the replacement of the cathode's material and radon contamination.}
\label{fig:oldULBP}
\end{figure}

\medskip
This replica was installed at the LSC lab (Fig. \ref{fig:set-up}, right), obtaining the same background level than in CAST (see table \ref{tab:BkgSummary}) but with a raw trigger rate of 0.2 Hz ( around 5 times lower than at surface). This result shows that cosmic muons are the most common event at surface, but do not dominate the background level after the offline analysis. In a second step, an extra 20 cm thick lead shielding was added, reducing the trigger rate to $\sim5\times10^{-3}$ Hz, and leading to a background level of 2$\times10^{-7}$\ckcs. This value was reproduced several times (Fig. \ref{fig:oldULBP}) and was quite independent on the electronic parameters, detector gain and analysis version \cite{alfredotesis}. We can conclude that the gamma flux is the main contribution to the final background in an unshielded detector.
 
At this ultra-low level achieved in the underground setup, the effects of other sources of background could be quantified experimentally: the aluminium cathode and the aire-borne Radon. The first one gives a contribution of $(5.2\pm1.2)\times10^{-7}$\cskc, which led to the replacement of CAST cathodes by radiopure copper ones. In addition, air-borne radon entered several times the detector setup due to interruptions in the nitrogen flux. Apart from an increase in background, some of them produced sparks at the Micromegas detector and an operational stop. In a controlled interruption, a value of $(3.0\pm0.8)\times10^{-9}\,$keV$^{-1}\,$cm$^{-2}\,$s$^{-1}$ per Bq/m$^{3}$ of air-borne Radon in the surrounding atmosphere was quantified by monitoring its concentration inside the Faraday cage with an alphaGUARD detector\cite{alphaguard}.

\begin{figure}[htb!]
\centering
\includegraphics[width=0.6\textwidth]{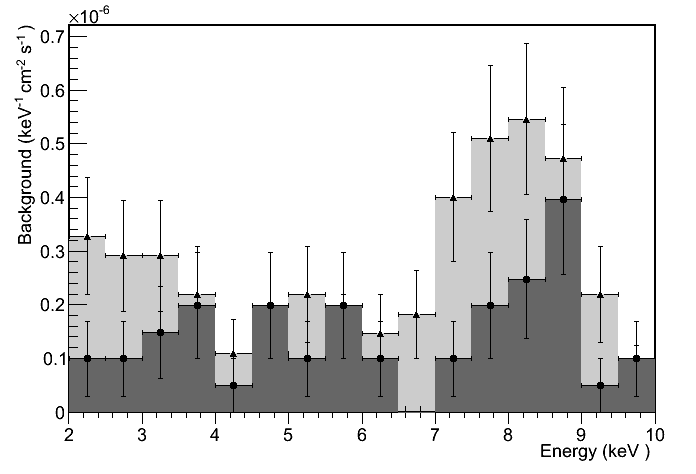}
\caption{Background registered by the CAST Micromegas detector with 10 cm lead thickness at the LSC using a copper cathode (light grey) and after the replacement of other chamber components with other more radiopure (dark grey).}
\label{fig:ULBspectra}
\end{figure}

\medskip
Most of the detector components not particularly radiopure were replaced, like the brass gas connectors by copper ones, leading to a low limit closer to $10^{-7}$\ckcs, as shown in Fig. ~\ref{fig:ULBspectra}. This level may be attributed to the intrinsic radioactivity of the detector components or of the inmediate surroundings (the front-end electronics is still poorly shielded for the detector in this setup). In any case, the lowest background level obtained underground is around 50 times better that in CAST.

\medskip
In a surface lab, the CAST-like setup was covered with a 20 cm thickness polyethylene shielding without noticeable influence in the background. It was concluded that neutrons were not likely to be an important background source at those levels. 

\section{The upgrade of CAST detectors in 2012. Present state-of-the-art.}\label{sec:sunset2012}

The experimental and simulation results described in the last two sections have been the motivation and basis for the last improvements in the shielding of the two CAST sunset Micromegas, just before the 2012 data taking campaign. The upgrade focused on reducing the dominant contributions of the external gamma flux and in particular the steel fluorescences produced by it. The lead thickness was increased to about 10 cm and the general shielding design improved to close the remaining open solid angles. In addition, the radiopurity of inner components was improved to allow for further potential background reductions.

\begin{figure}[htb!]
\centering
\includegraphics[height=0.27\textwidth]{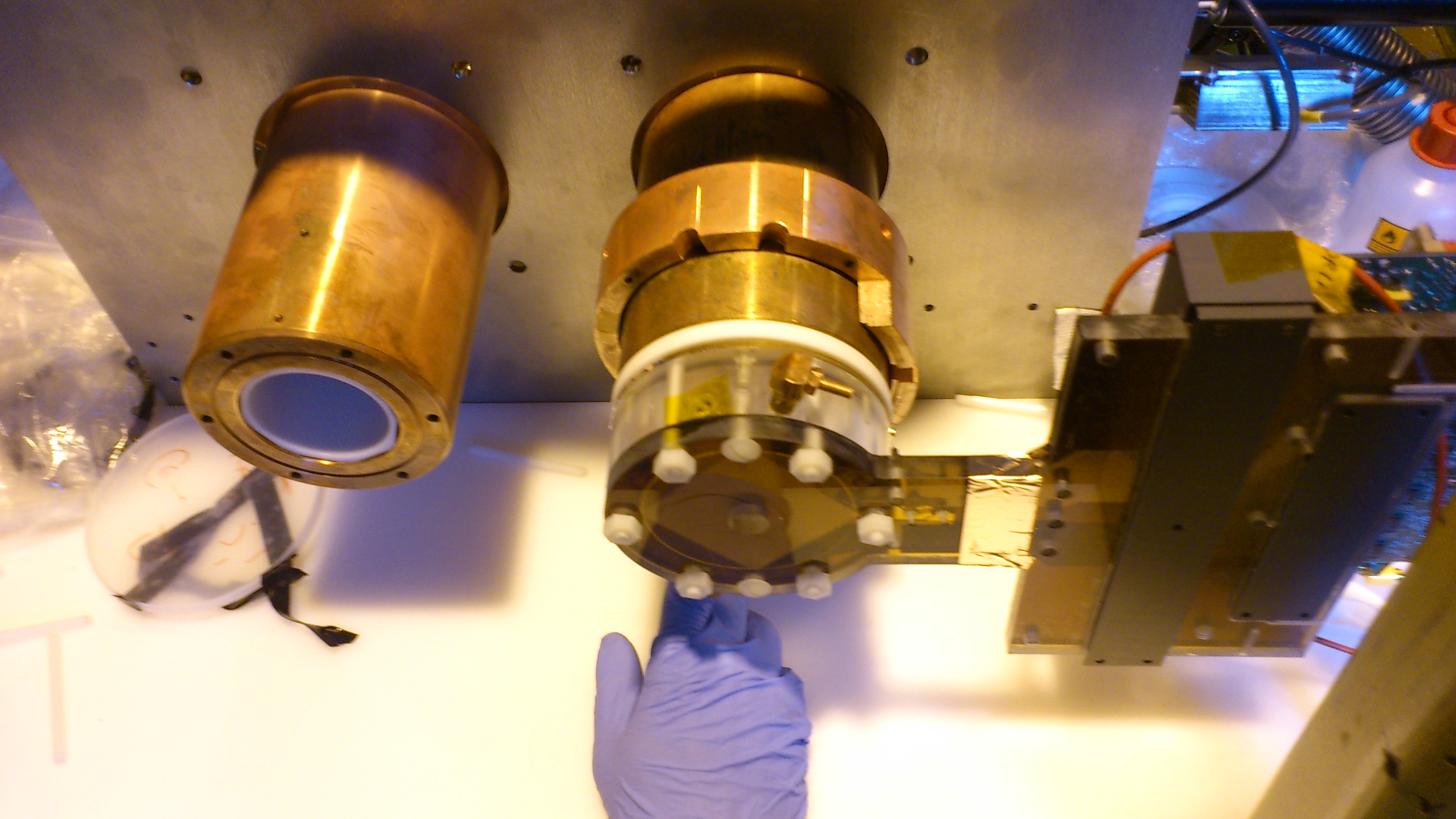}
\includegraphics[height=0.27\textwidth]{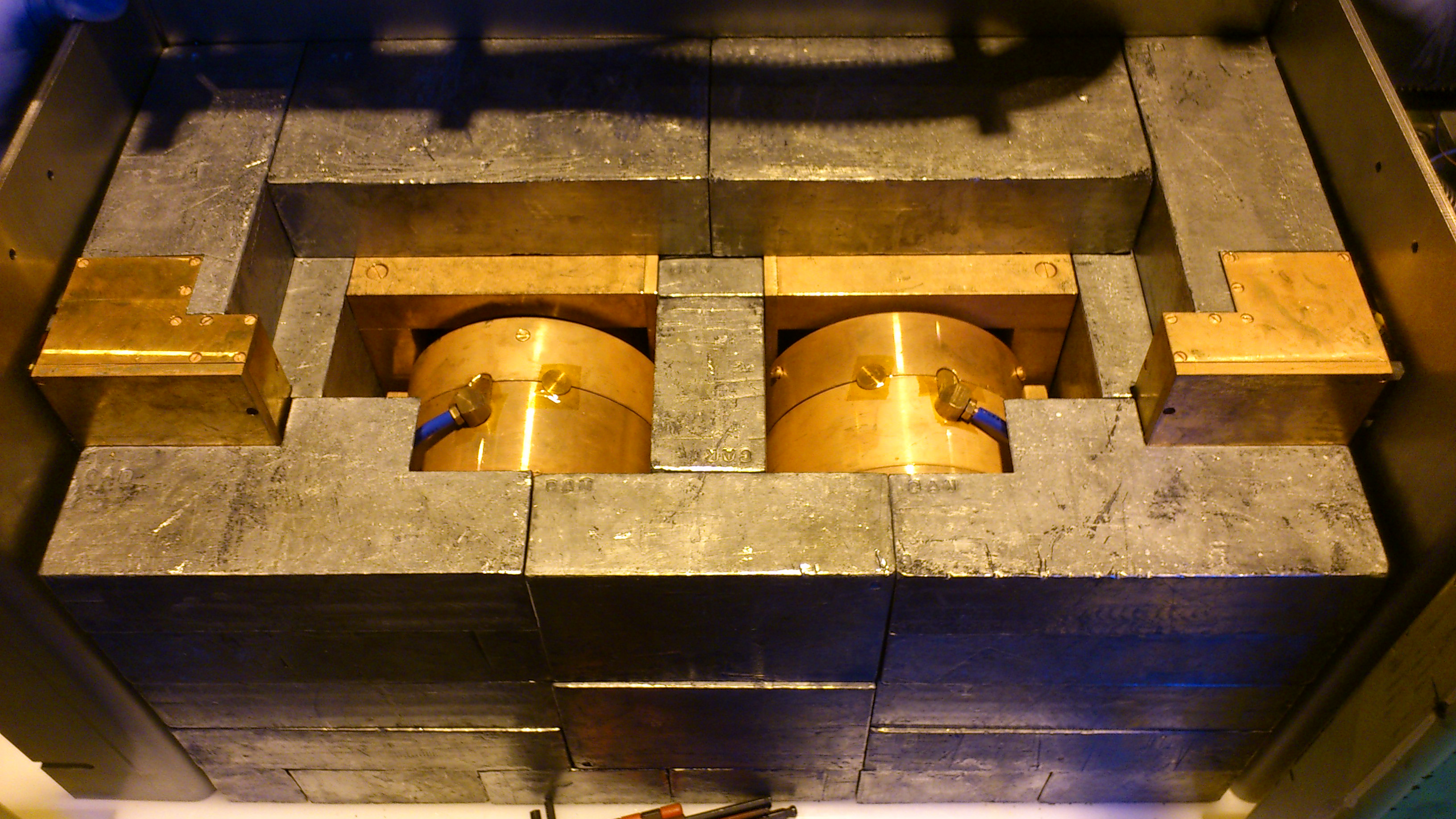}
\caption{Left: early step during sunset 2012 installation. The copper pipe with an inner Teflon coating is visible in the left bore; in the right bore it is shown the detector installation using Teflon and Delrin screws, nuts and gaskets, the new copper connections for the gas and the top piece of the new inner copper shielding which fits the pipe. Right: latter stage of the mounting showing the careful design to adjust copper shielding to detectors (closing as leak-tight as possible every opening), the Faraday cage and the lead layer. Note that not only detectors but also the pipe are shielded.}
\label{fig:2012shielding}
\end{figure}

\medskip
The current sunset setup (partially built) is shown in Fig.~\ref{fig:2012shielding}. The innermost copper shielding layer thickness is increased from 0.5 cm to 1 cm. Such an inner shielding is able to attenuate the Pb fluorescences ranging from 73 to 87 keV by a factor higher than $10^{3}$, as well as the 46.5 keV gamma from $^{210}$Pb decay.


\medskip
The new design reduces the remaining unshielded solid angle, like for example in the racket of the detector where the signals go out of the shielding sideways to the electronics. Also, the connection to the magnet bore is now done via a 18-cm long copper pipe that is also shielded with lead. This in addition improves and suppresses the steel fluorescences. This pipe has an inner teflon coating of 2.5 mm thick in order to shield 8 keV copper fluorescence by a factor higher than $10^{3}$.

\begin{figure}[htb!]
\centering
\includegraphics[width=0.95\textwidth]{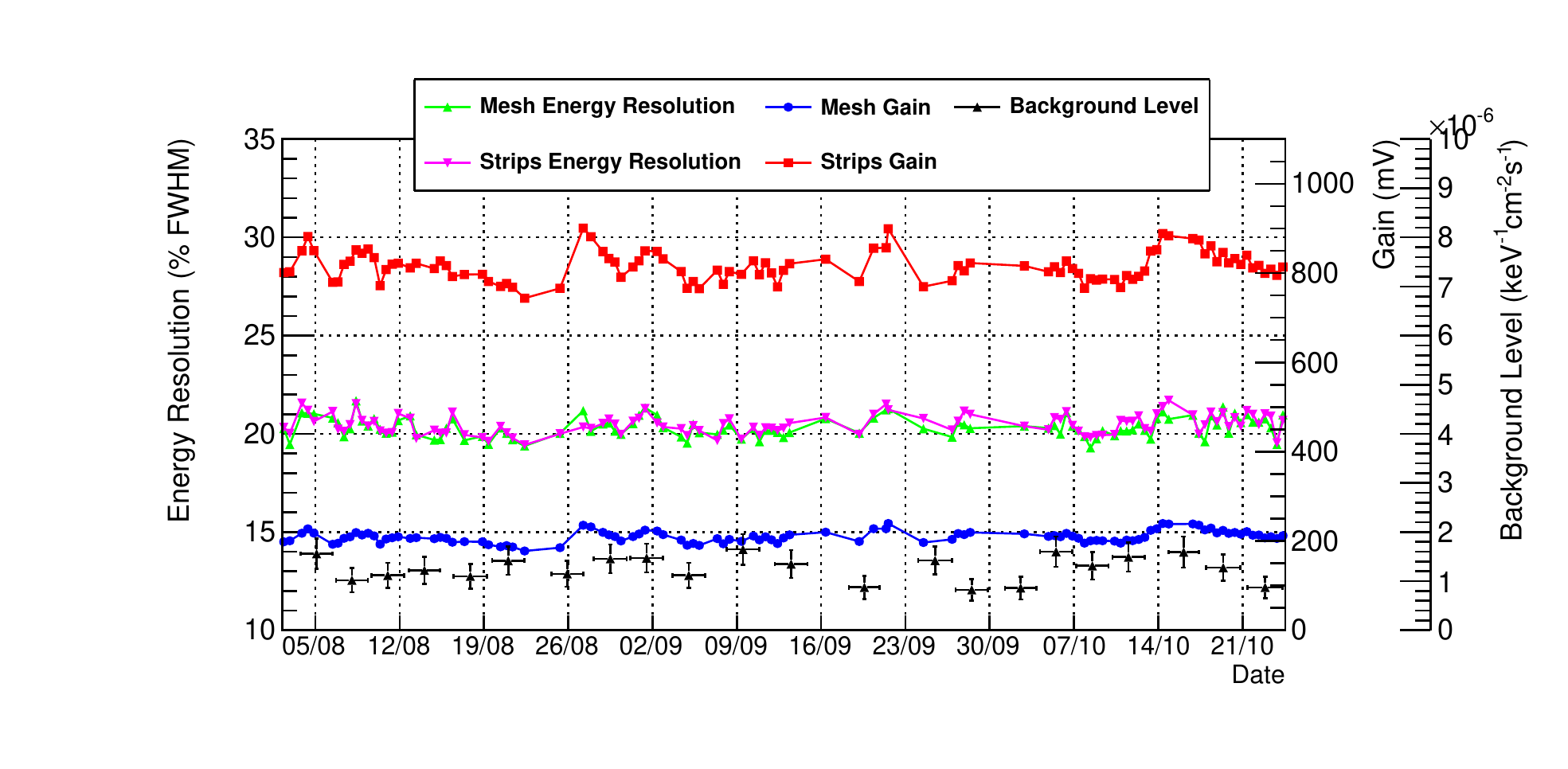}
\caption{Time evolution of  background level, gain and energy resolution of CAST Sunset 1 Micromegas detector during the 2012 data taking. The sharp change in the strips gain is related with an increase in the Micromegas voltage, while for the the mesh gain it is balanced with the Timing Amplifier settings.}
\label{fig:2012evolution}
\end{figure}

\medskip
In Fig.~\ref{fig:2012evolution} the performance of one of the sunset detectors during the 2012 data-taking is summarized. The new background levels are 1.3 (1.7)$\times10^{-6}$\cskc for sunset detector 1 (2)~\cite{ThPapaevangelou:SPSC2012}. That is about 4 times lower than that of the previous state-of-the-art setup, still operative in the sunrise side (see Fig.~\ref{fig:EvolCASTSunrise}). A direct comparison between both background spectra (see Fig.~\ref{fig:spectraEvolution}) shows that the main improvement, as expected, is the suppression of the stainless steel fluorescence peaks.

\section{Status of CAST Micromegas background model. }  \label{subsec:comparing}

\begin{figure}[htb!]
\centering
\includegraphics[height=0.32\textwidth]{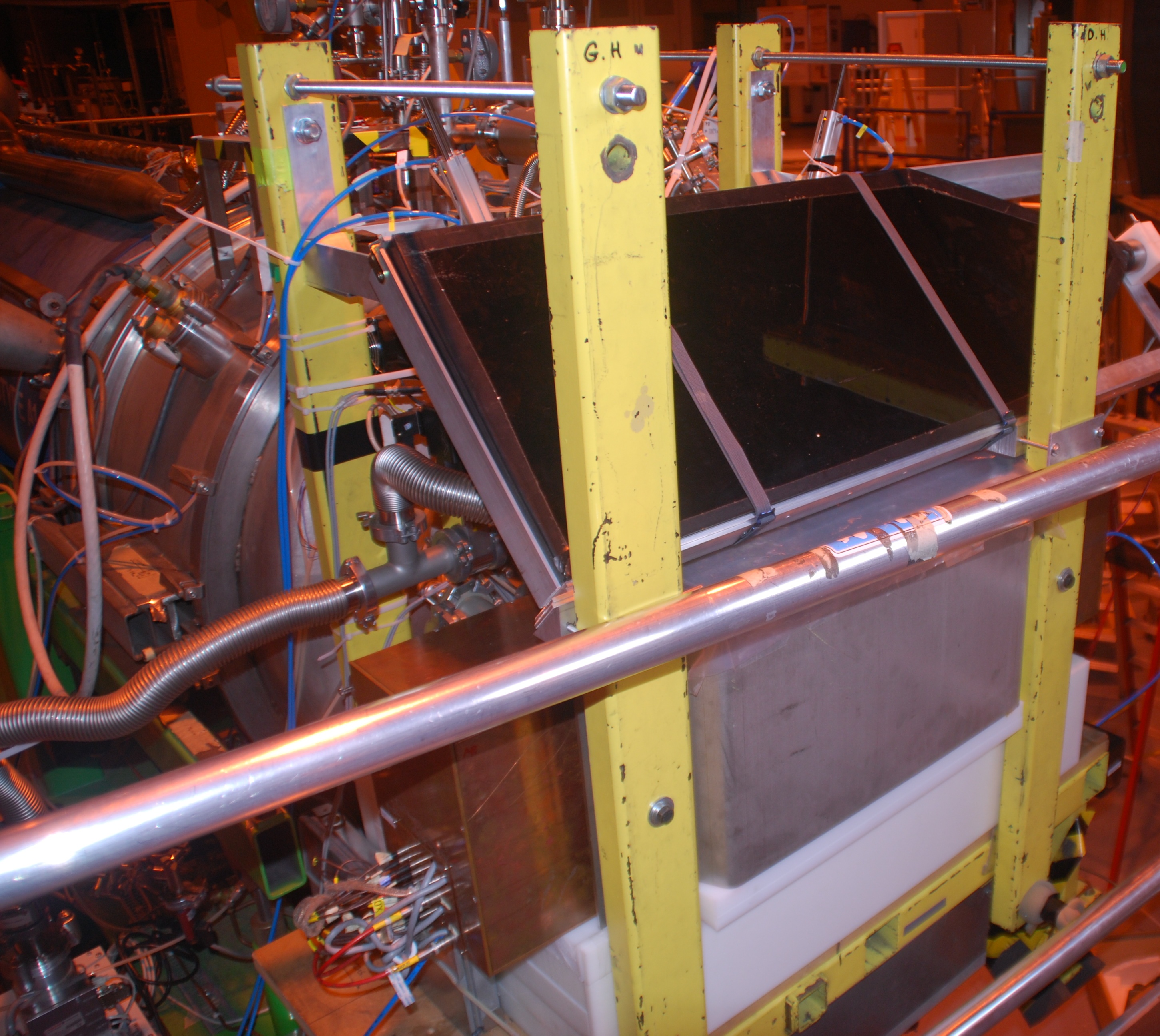}
\includegraphics[height=0.35\textwidth]{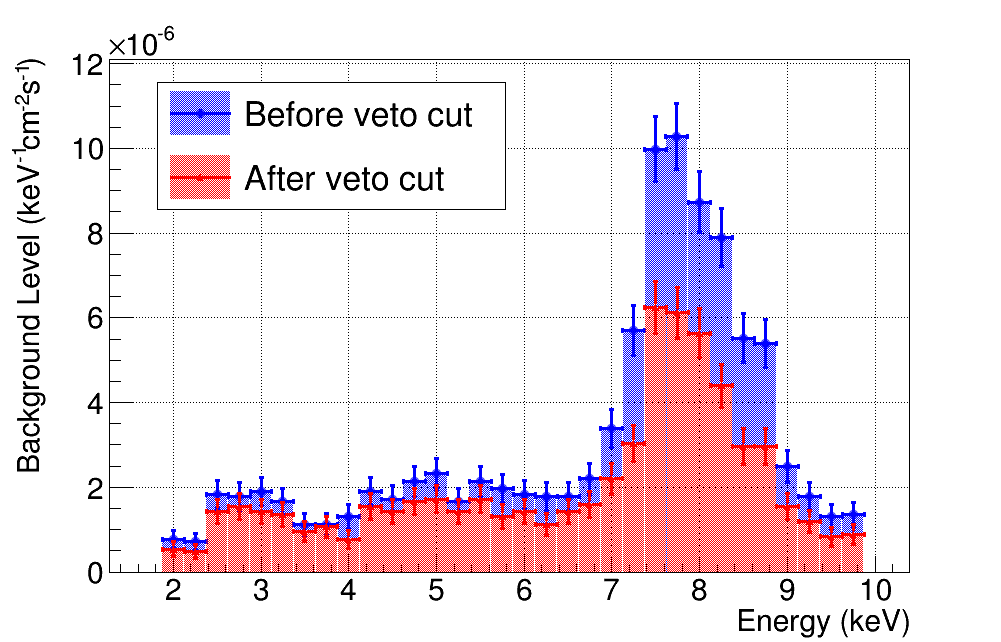}
\caption{Left: current installation of a cosmic muons veto in the sunset Micromegas setup. Right: sunset Micromegas background accumulated during the summer 2012 data taking campaign by detector 1, before and after applying the cosmic rays veto.}
\label{fig:vetoEffect}
\end{figure}

There is still a difference of one order of magnitude between background levels obtained in surface at CAST (shown in last section) and underground at the LSC (shown in section \ref{sec:tests}) with similar shielding . This fact suggests that cosmic muons are now becoming a substantial fraction of the remaining background. To evaluate experimentally this hypothesis, a preliminary test was done in the CAST 2012 sunset setup with a cosmic veto in anti-coincidence with the Micromegas DAQ (see Fig. \ref{fig:vetoEffect}, left). The geometrical coverage of the veto was only 44\% due to mechanical constraints of the setup, and the background was reduced by 25\%, as shown on the right of Fig. \ref{fig:vetoEffect}.


\begin{figure}[htb!]
\centering
\includegraphics[width=0.7\textwidth]{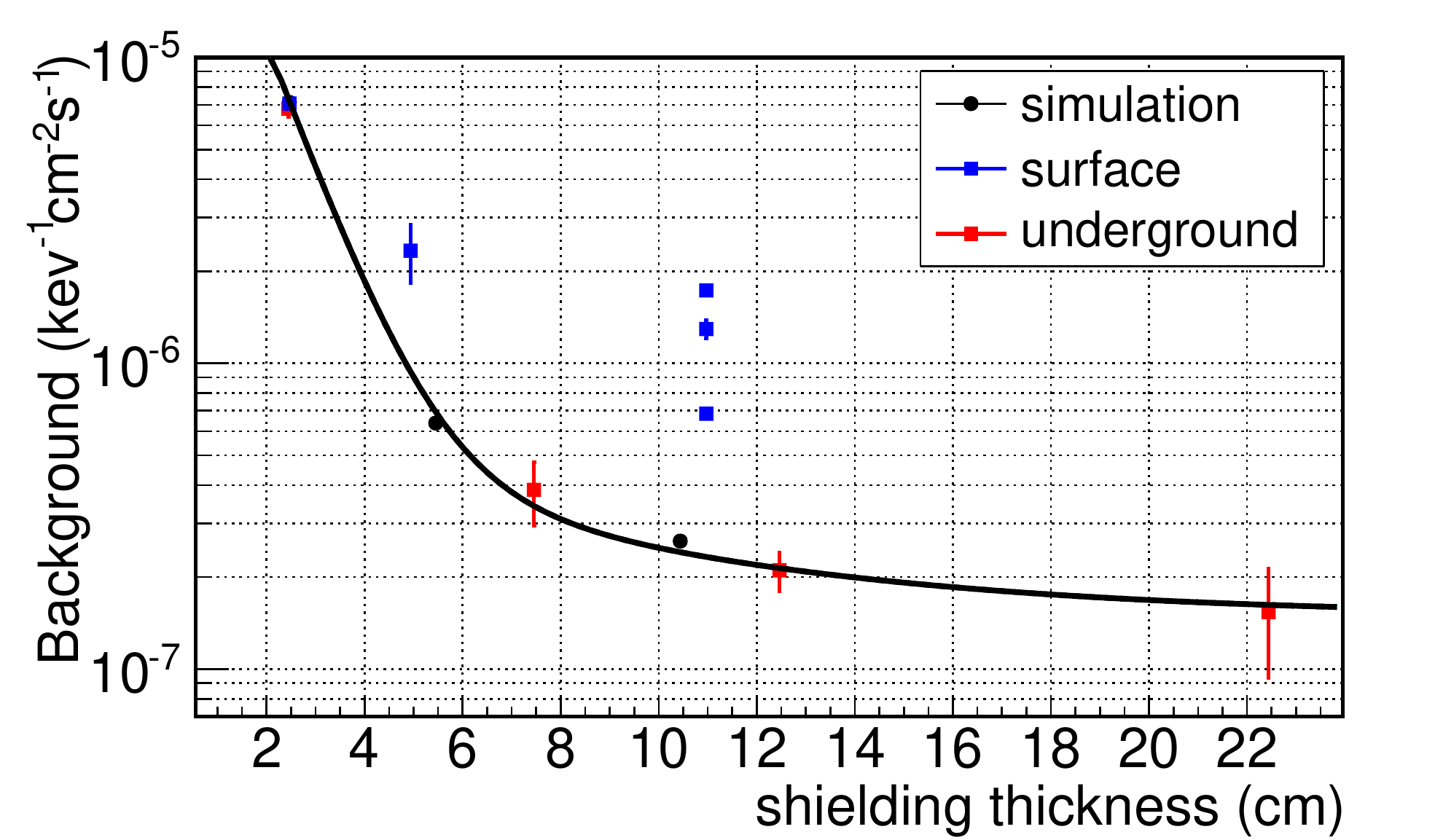}
\caption{Comparison between experimental background levels obtained at surface and underground for CAST detectors shielded with different lead thickness. Black circles and their corresponding fit represent the Monte Carlo prediction from the environmental gamma flux over an intrinsic radioactivity level.}
\label{fig:comparison-surface-underground}
\end{figure}

\begin{figure}[htb!]
\centering
\includegraphics[width=0.7\textwidth]{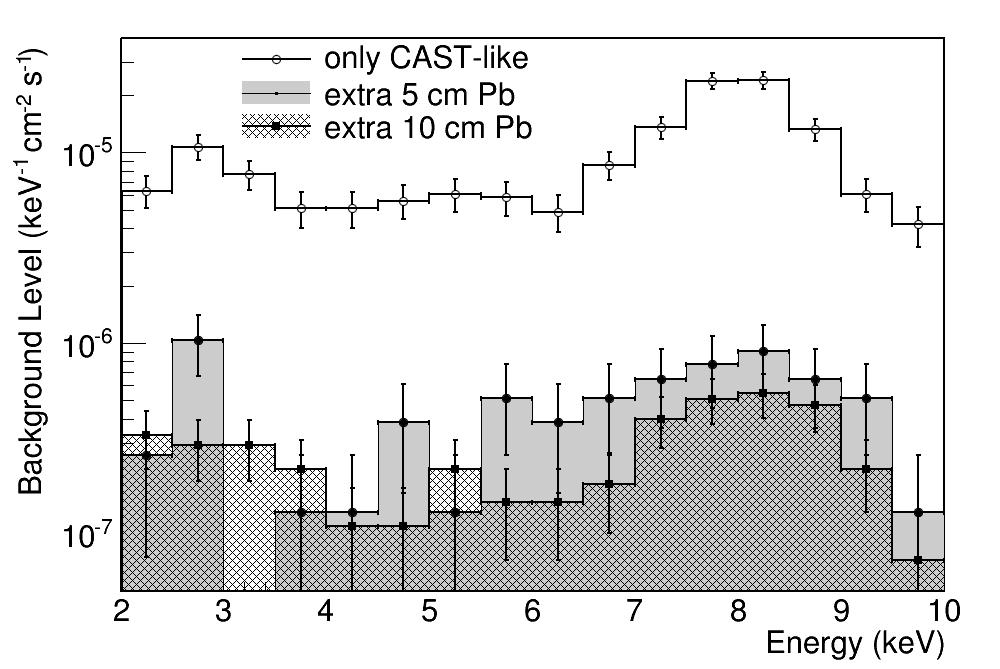}
\caption{Evolution of the M17 detector's background spectra with progressive shielding upgrades done undergrount at the LSC. The corresponding background levels can be found at the table 1.}
\label{fig:newShieldingUpgrade}
\end{figure}

In Fig. ~\ref{fig:comparison-surface-underground} we summarize the results in terms of background level as a function of shielding thickness for experimental tests (underground and at surface level) and simulations.  It is evident that a thicker lead shielding yields much more reduction of background in underground (red squares) than at surface (blue triangles). The simulation results (black dots) represent the contribution of environmental gamma flux. These values have been rescaled so as to make the first experimental background point with the lightest shielding match its simulated value, presented in section~\ref{sec:sims}. Later, the experimental background contribution by intrinsic radioactivity of the detectors (presented in section~\ref{sec:tests}) has been added to the simulated values. The fit to the simulated results matches quite well with the experimental values. This fact confirms the main contribution to background of gammas at underground tests. The background energy spectra of the underground measurements are shown in Fig. \ref{fig:newShieldingUpgrade}.

\medskip
As shown in Fig.~\ref{fig:comparison-surface-underground}, the environmental gamma flux and the intrinsic radioactivity only cannot account for the background progression at surface. The three measurements carried out at surface with 11 cm thickness correspond from highest to lowest level to setups with no muon veto, $\sim44$\% and $\sim75$\% efficient geometric coverage respectively. The two first points correspond to a CAST Sunset detector during the 2012 data taking campaign, while the latest correspond to a replica of that setup operated at surface in the University of Zaragoza. An increase in the veto efficiency leads to background levels closer to the underground level for a given shielding thickness. A more efficient veto is foreseen in the next shielding upgrade for the forthcoming CAST data taking.

The total contribution of muons to the background level is around $\sim 2\times10^{-6}\,$\ckcs. It was also found that the dependence of this contribution with the shielding thickness is negligible at least up to 10 cm of lead, which indicates that secondary particles induced in the shielding are not contributing significantly. Therefore, muon-induced events are mainly due to fluorescences created by the primary particles close to the detection medium, and so, easy to be tagged with a veto.

\section{Future prospects.}\label{sec:prospects}

The long term goal of this R\&D is to satisfy the background requirements of IAXO, i.e. a level of $10^{-7}$\ckcs or lower at surface level. The present picture of CAST background model suggests that the strategies followed until now are not yet exhausted and future improvements are expected. We will briefly mention the current lines of work:


\subsection*{Detector's design}
An improved detector, based on a redesign of the readout and chamber's body will be installed and operated in the CAST sunrise setup for the 2013 data taking campagin. In the new design~\cite{ThPapaevangelou:SPSC2012}, that will be the subject of a future publication, the gas chamber, Faraday Cage and inner shielding are completely integrated. All the high voltages are driven along the detector printed board, in this way the connectors to the electronics, being potentially radioactive, are completely moved out of the whole copper plus lead shielding. The detector's x-ray window and the pipe that connects it to the magnet will have a smaller diameter, as the detector will operate with an x-ray focussing device for the 2014 campaign. Therefore the main limitations of the basic CAST-like setup will be overcome and the intrinsic limit of the new setup (in the sense that was explained in section~\ref{sec:tests}) can be expected to be below $10^{-7}\,$\ckcs. Finally, the drift field quality has been improved by means of a radiopure field shaper made of copper strips on a multilayer polymide flexible circuit.

On the other hand, the development of segmented mesh detectors~\cite{Thomas_SegmentedMesh} coupled to autotrigger electronics is very promising for its application to the rare event searches field. The segmentation of the mesh will provide simplification of the microbulk production and mass minimization, allowing to pave large surfaces with high radiopurity.

\subsection*{Shielding.}
As it was already explained, the implementation of a high efficient cosmic muon veto will be essential to achieve background levels in CAST similar to underground ones. The new scintillator to be installed in the CAST sunset docking point for the 2013 campaign has been computed to have a coverage for both Micromegas detectors of $\simeq95$\% of the cosmic muons. Depending on the effect of this upgrade, further improvements in the passive shielding (e.g. increasing the thickness) may be motivated.


\subsection*{Electronics.}
\begin{figure}[htb!]
\centering
\includegraphics[width=75mm]{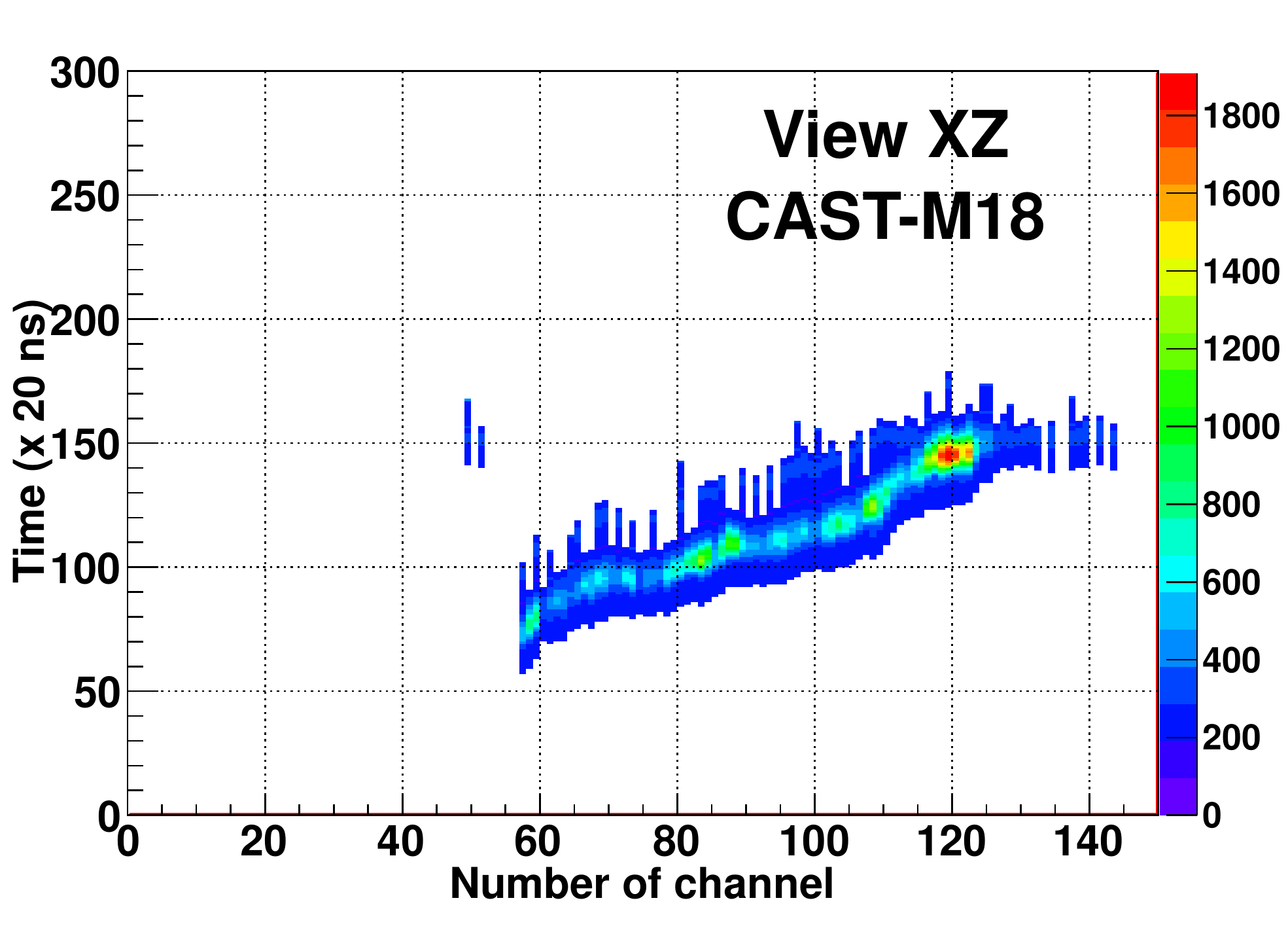}
\includegraphics[width=75mm]{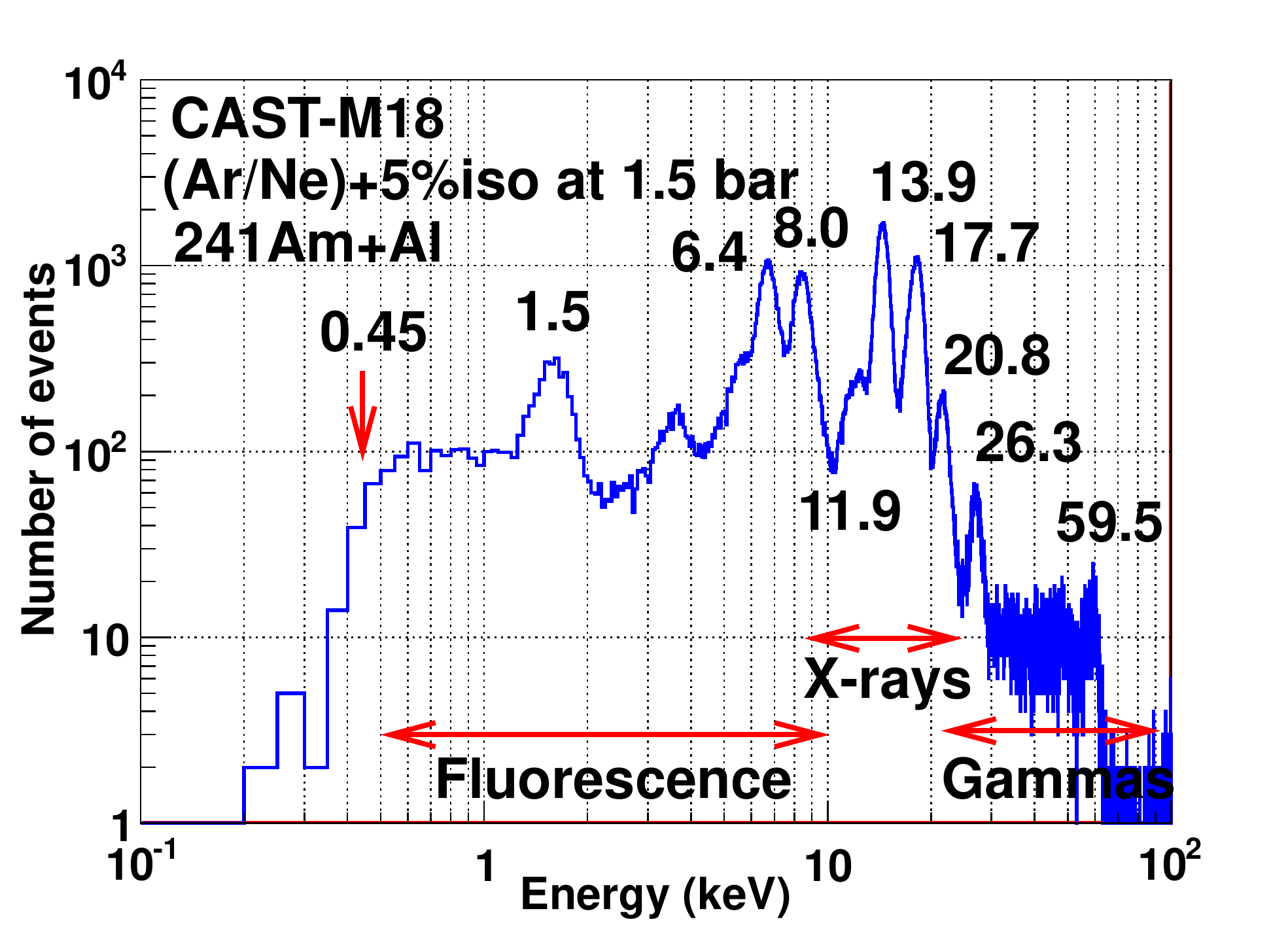}
\caption{Left: The 2D view of an electron acquired in a background run by a CAST Micromegas detector reading the strips with AFTER-based electronics. Right: Energy spectra generated by the strips of a CAST Micromegas detector read by the AFTER-based electronics when the detector was illuminated by a $^{241}$Am source covered by an aluminium foil. Apart from the gammas and x-rays emitted by the source, fluorescence of copper (8 keV), iron (6.4 keV) and aluminium (1.5 keV) are present. The energy threshold is situated around 450 eV.}
\label{fig:T2KElec}
\end{figure}

For the next CAST data-taking campaign, TPC-like electronics based on the AFTER chip~\cite{AFTER,19273929}, will replace the Gassiplex-based electronics. The advantage of these electronics is the time information for each strip, unlike the former one for which the only the integrated charge information is available. As an example, the 2D view of a background event acquired by a CAST Micromegas detector and AFTER-based electronics is shown on the left of Fig. \ref{fig:T2KElec}. As the signal to noise ratio is improved, it will also lower the threshold to sub-keV energies. First tests have shown that values as low as 450 eV can easily obtained with AFTER chip (see Fig. \ref{fig:T2KElec} right). If systems with autotrigger capabilities (like AGET~\cite{AGET}) are available, even lower energy thresholds could be achieved in the future.
 
\subsection*{Analysis.}

Since x-rays from steel fluorescences, that were one of the most important contribution to background in RoI, have been avoided, the remaining events may be susceptible to be rejected with an improved detector and/or analysis performance that could follow the detector and acquisition upgrades. Some reasons to expect that could be:

\begin{itemize}

\item The upgrade of the chamber structure of the new detectors will provide a more uniform drift field. This fact will improve the detector homogeneity and the analysis performance.

\item The electronics upgrade will give the time information of each strip, opening new possibilities for background rejection and an enhancement of the signal efficiency at low energies due to the improved signal-to-noise ratio.


\item The energy dependence of analysis observables is an open field for optimization. In this direction, an electron beam based on PIXE (Particle Induced x-rays Emission) has been installed and tested at CAST detector laboratory at CERN~\cite{ThPapaevangelou:SPSC2012}. This beam generates different fluorescence lines in the RoI depending on the target material like Al (1.49 keV), Ti (4.41 keV) or Cu (8.9 keV). These lines are complementary to the ones of usual CAST calibration runs at 3 and 5.9 keV.

\end{itemize}

\section{Conclusions.} \label{sec:conclusions}

Small gas TPCs read by Micromegas planes are an optimal technological choice for low background x-ray detection. They have been used and improved within the CAST experiment at CERN for about a decade, and more recently are being actively developed within the T-REX R\&D project. We have described the latest status of CAST Micromegas background model, based on different underground/surface tests with different shielding configurations and on Monte Carlo simulations. 

\begin{figure}[htb!]
\centering
\includegraphics[height=25pc]{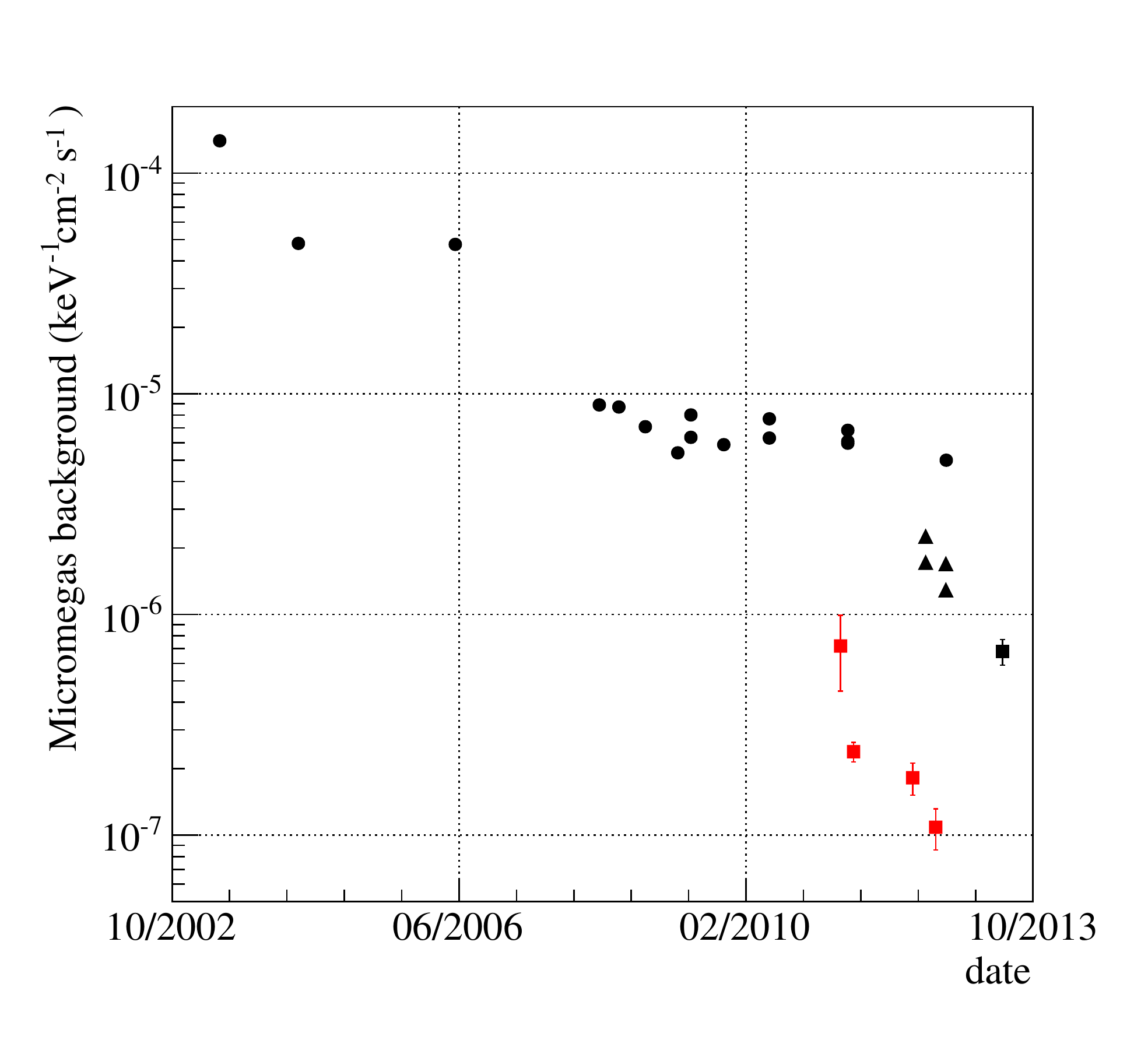}
\caption{Micromegas background evolution since the first detector installation in 2002. The black points correspond to the values obtained with the different detectors during true data-taking campaigns. The black squares correspond to the values obtained with the upgraded sunset setup, preliminarily and after application of the muon veto. The red series corresponds to the values obtained with heavy shielding configurations in the Canfranc Underground Laboratory, the last two points obtained only with only 10 cm lead thickness but improved detector radiopurity.}
\label{fig:histoPlot}
\end{figure}

Figure \ref{fig:histoPlot} shows a compilation of background levels achieved by Micromegas detectors over the history of CAST. Since the first generation of detectors in 2002, the background has been reduced by about two orders of magnitude, and current best background levels are about 4--5 orders of magnitude below the raw level of the unshielded detector. This background reduction is achieved thanks to the combined effect of:

\begin{itemize}
\item detector active and passive shielding.

\item simple detector's geometry with component materials optimized for radiopurity.

\item high granularity of the readout, providing a high-quality topological information of the event ionization in the conversion gas, on which to apply offline discrimination algorithms.
\end{itemize}

\begin{figure}[htb!]
\centering
\includegraphics[width = 0.7\textwidth, angle = 0]{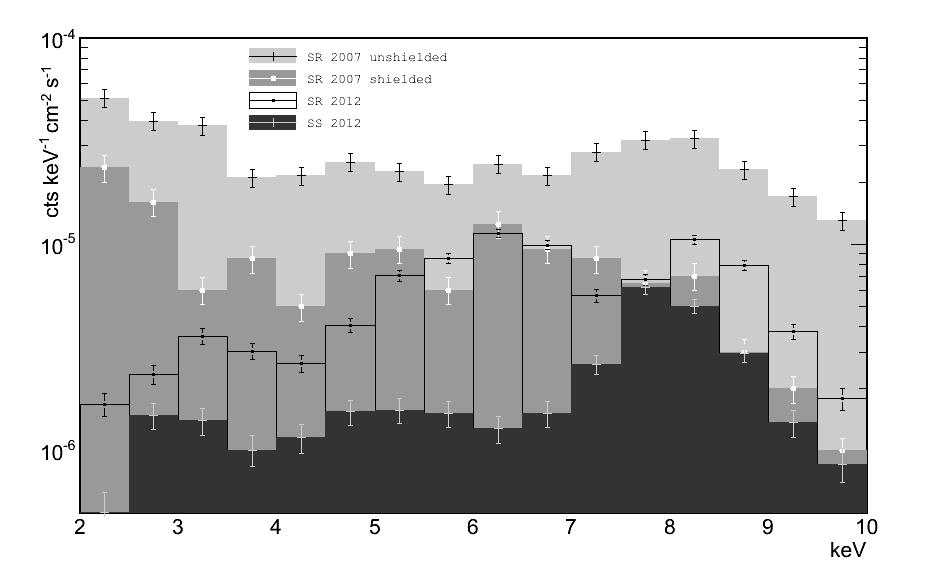}
\caption{Evolution of CAST Micromegas background spectra since 2007. The spectra shown correspond to: unshielded classical Micromegas (\emph{SR 2007 unshielded}); the first light shielding (2.5 cm thickness of lead, as described in section 3.3) of the same detector (\emph{SR 2007 shielded}); the 2012 CAST sunrise detector, with no change in the setup shielding (\emph{SR 2012}); the 2012 CAST sunset detector after application of the cosmic veto cut (\emph{SS 2012}).}
\label{fig:spectraEvolution}
\end{figure}

The first remarkable improvement (a factor of 3) was due to the addition of a 2.5 cm thick lead shielding in 2007~\cite{Aune:2009zz}. During 2008--11 period, the background level slightly decreased due to two related causes: the consolidation of the microbulk technique and the refinement of discrimination routines. That fact is clearly appreciated at the spectra comparison (plotted in Fig. ~\ref{fig:spectraEvolution}) between classical (2007) and microbulk (2012) micromegas taken in the same shielding configuration.

The Sunset 2012 upgrade (described in section~\ref{sec:sunset2012})  produced an extra reduction factor of 4, and the implementation of a cosmic muon veto produced a further background reduction of 25\%. This is illustrated by the two couples of triangled points plotted together in Fig.~\ref{fig:histoPlot}), that represent the values of both CAST sunset detectors before and after the application of the cosmic muon veto. They  anticipated a new improvement due to the introduction of a higher efficient muon veto ($\sim75$\%) in a test setup, as it is shown by the last squared black point in the figure. This fact motivates the specific production of scintillator vetos for the CAST 2013 data-taking campaign, which will reach a coverage higher than 90\%. 

To a lesser extent, another source of background at surface may be the external gamma radiation that still penetrates the current shielding design by the opening solid angle of the pipe connecting the detector to the magnet. This insight is the basis for the newly designed detector that will be fullly installed with x-ray optics in 2014, when it is expected to reach background levels close to $\sim10^{-7}$ \ckcs. The details of this new design and the experience obtained with it will be object of a future publication.

\medskip
 In parallel, the underground setup at LSC setup has proven background levels down to $\sim10^{-7}$ \ckcs achieved by means of replacing some detector components by other more radiopure. We attribute this limit to the current internal radiopurity budget of the setup. The effectiveness of the inner shielding (inside the Faraday cage) to electronics is now being studied, and there are prospects to reach even lower levels of background.
 
\medskip
These detectors are the most promising technology to be part of IAXO, a new generation axion helioscope. IAXO aims to improve substantially the present sensitivity to solar axions (that of CAST), by means of a large superconducting magnet, large x-ray optics and very low background x-ray detectors. The sensitivity prospects of IAXO rely, among others, on very stringent requirements for the x-ray detectors. Background levels of at least $\sim10^{-7}$ \cskc and down to $\sim10^{-8}$ \cskc are one of the experimental parameters required by the IAXO proposal~\cite{Irastorza:1567109}. Micromegas detectors like the ones here developed have realistic prospects to fulfil these requirements. The developments associated with low-background Micromegas detectors have interest for other rare event searches, like Dark Matter WIMP detectors or double beta decay.

\section*{Acknowledgements}
We want to thank our colleagues of CAST for many years of collaborative work in the experiment, and many helpful discussions and encouragement. We thank R. de Oliveira and his team at CERN for the manufacturing of the microbulk readouts. We also thank the LSC staff for their help in the support of the Micromegas setup at the LSC; Authors would like to acknowledge the use of Servicio General de Apoyo a la Investigaci\'on-SAI, Universidad de Zaragoza. F.I. acknowledges the support of the Eurotalents program. We acknowledge support from the European Commission under the European Research Council T-REX Starting Grant ref. ERC-2009-StG-240054 of the IDEAS program of the 7th EU Framework Program. We also acknowledge support from the Spanish Ministry of Economy and Competitiveness (MINECO) under contracts ref. FPA2008-03456 and FPA2011-24058, as well as under the CPAN project ref. CSD2007-00042 from the Consolider-Ingenio 2010 program. Part of these grants are funded by the European Regional Development Fund (ERDF/FEDER).

\bibliographystyle{JHEP}

\begin{thebibliography}{10}

\bibitem{Peccei:1977hh}
R.~D. Peccei and H.~R. Quinn, {\it {CP conservation in the Presence of
  Instantons}},  {\em Phys. Rev. Lett.} {\bf 38} (1977) 1440--1443.

\bibitem{Cheng:1987gp}
H.-Y. Cheng, {\it {The Strong CP Problem Revisited}},  {\em Phys.Rept.} {\bf
  158} (1988) 1.

\bibitem{Ringwald:2012cu}
A.~Ringwald, {\it {Searching for axions and ALPs from string theory}},
  \href{http://xxx.lanl.gov/abs/1209.2299}{{\tt arXiv:1209.2299}}.

\bibitem{Abbott:1982af}
L.~Abbott and P.~Sikivie, {\it {A Cosmological Bound on the Invisible Axion}},
  {\em Phys.Lett.} {\bf B120} (1983) 133--136.

\bibitem{Dine:1982ah}
M.~Dine and W.~Fischler, {\it {The Not So Harmless Axion}},  {\em Phys.Lett.}
  {\bf B120} (1983) 137--141.

\bibitem{Preskill:1982cy}
J.~Preskill, M.~B. Wise, and F.~Wilczek, {\it {Cosmology of the Invisible
  Axion}},  {\em Phys.Lett.} {\bf B120} (1983) 127--132.

\bibitem{Sikivie:1983ip}
P.~Sikivie, {\it {Experimental tests of the invisible axion}},  {\em Phys. Rev.
  Lett.} {\bf 51} (1983) 1415.

\bibitem{Zioutas:2004hi}
{\bf CAST} Collaboration, K.~Zioutas et~al., {\it {First results from the CERN
  Axion Solar Telescope (CAST)}},  {\em Phys. Rev. Lett.} {\bf 94} (2005)
  121301, [\href{http://xxx.lanl.gov/abs/hep-ex/0411033}{{\tt
  hep-ex/0411033}}].

\bibitem{kuster2007}
M.~Kuster, H.~Braeuninger, S.~Cebri{\'a}n, M.~Davenport, C.~Eleftheriadis,
  J.~Englhauser, H.~Fischer, J.~Franz, P.~Friedrich, R.~Hartmann, F.~H.
  Heinsius, D.~H.~H. Hoffmann, G.~Hoffmeister, J.~N. Joux, D.~Kang,
  K.~Koenigsmann, R.~Kotthaus, T.~Papaevangelou, C.~Lasseur, A.~Lippitsch,
  G.~Lutz, J.~Morales, A.~Rodriguez, L.~Strueder, J.~Vogel, and K.~Zioutas,
  {\it {The x-ray telescope of CAST}},  {\em {NEW JOURNAL OF PHYSICS}} {\bf
  {9}} ({JUN 22}, {2007}).

\bibitem{ThPapaevangelou:SPSC2012}
T.~Papaevangelou, {\it {2012 Status report of the CAST Experiment and Running
  in 2013-2014}},  Tech. Rep. CERN-SPSC-2012-028, CERN, Geneva, Oct, 2012.

\bibitem{unser_cast_paper}
C.~Krieger, J.~Kaminski, and K.~Desch, {\it {InGrid-based X-ray detector for
  low background searches}}, . Submitted to Nucl. Instr. and Meth.

\bibitem{proc_ingrid}
J.~Kaminski, C.~Krieger, Y.~Bilevych, T.~Krautscheid, and K.~Desch, {\it
  {GridPix as a candidate for the future of CAST}}, . Submitted to the
  conference proceedings of the Sixth Symposium on large TPCs for Low Energy
  Rare Event Detection', 2012, Paris, published in JPCS.

\bibitem{Andriamonje:2007ew}
{\bf CAST} Collaboration, S.~Andriamonje et~al., {\it {An improved limit on the
  axion-photon coupling from the CAST experiment}},  {\em JCAP} {\bf 0704}
  (2007) 010, [\href{http://xxx.lanl.gov/abs/hep-ex/0702006}{{\tt
  hep-ex/0702006}}].

\bibitem{Arik:2008mq}
{\bf CAST} Collaboration, E.~Arik et~al., {\it {Probing eV-scale axions with
  CAST}},  {\em JCAP} {\bf 0902} (2009) 008,
  [\href{http://xxx.lanl.gov/abs/0810.4482}{{\tt arXiv:0810.4482}}].

\bibitem{Aune:2011rx}
{\bf CAST} Collaboration, E.~Arik et~al., {\it Search for sub-ev mass solar
  axions by the cern axion solar telescope with $^{3}\mathrm{He}$ buffer gas},
  {\em Phys. Rev. Lett.} {\bf 107} (Dec, 2011) 261302.

\bibitem{Arik:2013nya}
M.~Arik, S.~Aune, K.~Barth, A.~Belov, S.~Borghi, et~al., {\it {CAST solar axion
  search with $^3$He buffer gas: Closing the hot dark matter gap}},
  \href{http://xxx.lanl.gov/abs/1307.1985}{{\tt arXiv:1307.1985}}.

\bibitem{Irastorza:2011gs}
I.~G. Irastorza, F.~Avignone, S.~Caspi, J.~Carmona, T.~Dafni, et~al., {\it
  {Towards a new generation axion helioscope}},  {\em JCAP} {\bf 1106} (2011)
  013, [\href{http://xxx.lanl.gov/abs/1103.5334}{{\tt arXiv:1103.5334}}].

\bibitem{Irastorza:1567109}
I.~G. Irastorza, {\it The international axion observatory iaxo. letter of
  intent to the cern sps committee.},  Tech. Rep. CERN-SPSC-2013-022.
  SPSC-I-242, CERN, Geneva, Aug, 2013.

\bibitem{Shilon:2012te}
I.~Shilon, A.~Dudarev, H.~Silva, and H.~Kate, {\it {Conceptual Design of a New
  Large Superconducting Toroid for IAXO, the New International AXion
  Observatory}},  {\em IEEE Trans. Appl. Supercond.} {\bf 23} (2012)
  [\href{http://xxx.lanl.gov/abs/1212.4633}{{\tt arXiv:1212.4633}}].

\bibitem{Irastorza2011_EAS}
I.~Irastorza, J.~Castel, S.~Cebri{\'a}n, T.~Dafni, G.~Fanourakis,
  E.~Ferrer-Ribas, D.~Fortuño, L.~Esteban, J.~Gal{\'a}n, J.~Garc{\'i}a,
  A.~Gardikiotis, J.~Garza, T.~Geralis, I.~Giomataris, H.~G{\'o}mez,
  D.~Herrera, F.~Iguaz, G.~Luz{\'o}n, J.~Mols, A.~Ortiz, T.~Papaevangelou,
  A.~Rodr{\'i}guez, J.~Ruz, L.~Segui, A.~Tom{\'a}s, T.~Vafeiadis, and
  S.~Yildiz, {\it Status of {R\&D} on micromegas for rare event searches : The
  {T-REX} project},  {\em EAS Publications Series} {\bf 53} (0, 2012) 147--154.

\bibitem{Ahlen:2009ev}
S.~Ahlen et~al., {\it {The case for a directional dark matter detector and the
  status of current experimental efforts}},  {\em Int. J. Mod. Phys.} {\bf A25}
  (2010) 1--51, [\href{http://xxx.lanl.gov/abs/0911.0323}{{\tt
  arXiv:0911.0323}}].

\bibitem{Cebrian:2010nw}
S.~Cebri{\'a}n et~al., {\it {Micromegas readouts for double beta decay
  searches}},  {\em JCAP} {\bf 1010} (2010) 010,
  [\href{http://xxx.lanl.gov/abs/1009.1827}{{\tt arXiv:1009.1827}}].

\bibitem{Giomataris:1995fq}
Y.~Giomataris, P.~Rebourgeard, J.~P. Robert, and G.~Charpak, {\it {MICROMEGAS:
  A high-granularity position-sensitive gaseous detector for high particle-flux
  environments}},  {\em Nucl. Instrum. Meth.} {\bf A376} (1996) 29--35.

\bibitem{Andriamonje:2010zz}
S.~Andriamonje, D.~Attie, E.~Berthoumieux, M.~Calviani, P.~Colas, et~al., {\it
  {Development and performance of Microbulk Micromegas detectors}},  {\em
  JINST} {\bf 5} (2010) P02001.

\bibitem{Iguaz:2012ur}
F.~Iguaz, E.~Ferrer-Ribas, A.~Giganon, and I.~Giomataris, {\it
  {Characterization of microbulk detectors in argon- and neon-based mixtures}},
   {\em JINST} {\bf 7} (2012) P04007,
  [\href{http://xxx.lanl.gov/abs/1201.3012}{{\tt arXiv:1201.3012}}].

\bibitem{Dafni:2009jb}
T.~Dafni et~al., {\it {Energy resolution of alpha particles in a microbulk
  Micromegas detector at high pressure Argon and Xenon mixtures}},  {\em Nucl.
  Instrum. Meth.} {\bf A608} (2009) 259--266,
  [\href{http://xxx.lanl.gov/abs/0906.0534}{{\tt arXiv:0906.0534}}].

\bibitem{Cebrian:2012sp}
S.~Cebrian, T.~Dafni, E.~Ferrer-Ribas, I.~Giomataris, D.~Gonzalez-Diaz, et~al.,
  {\it {Micromegas-TPC operation at high pressure in xenon-trimethylamine
  mixtures}},  {\em JINST} {\bf 8} (2013) P01012,
  [\href{http://xxx.lanl.gov/abs/1210.3287}{{\tt arXiv:1210.3287}}].

\bibitem{Cebrian:2010ta}
S.~Cebri{\'a}n et~al., {\it {Radiopurity of Micromegas readout planes}},  {\em
  Astropart. Phys.} {\bf 34} (2011) 354--359,
  [\href{http://xxx.lanl.gov/abs/1005.2022}{{\tt arXiv:1005.2022}}].

\bibitem{Aune:2009zz}
{\bf CAST} Collaboration, S.~Aune et~al., {\it {New Micromegas detectors in the
  CAST experiment}},  {\em Nucl. Instrum. Meth.} {\bf A604} (2009) 15--19.

\bibitem{mmcast_mpgd09}
J.~Gal\'an, ``{Micromegas detectors in the CAST experiment}.''
\newblock talk given at the Workshop on Micropattern Gas Detectors, 12-17 June
  2009, Kolympari, Crete, Greece. http://candia.inp.demokritos.gr/mpgd2009/.

\bibitem{XRS:XRS1319}
A.~Tomas, S.~Aune, T.~Dafni, G.~Fanourakis, E.~Ferrer-Ribas, J.~Galan,
  A.~Gardikiotis, J.~A. Garcia, T.~Geralis, I.~Giomataris, H.~Gomez, F.~J.
  Iguaz, I.~G. Irastorza, G.~Luzon, J.~Morales, T.~Papaevangelou, A.~Rodriguez,
  J.~Ruz, L.~Segui, T.~Vafeiadis, and S.~C. Yildz, {\it The new micromegas
  x-ray detectors in cast},  {\em X-Ray Spectrometry} {\bf 40} (2011), no.~4
  240--246.

\bibitem{Bahcall:2004fg}
J.~N. Bahcall and M.~Pinsonneault, {\it {What do we (not) know theoretically
  about solar neutrino fluxes?}},  {\em Phys.Rev.Lett.} {\bf 92} (2004) 121301,
  [\href{http://xxx.lanl.gov/abs/astro-ph/0402114}{{\tt astro-ph/0402114}}].

\bibitem{Santiard:1994ps}
J.~Santiard, W.~Beusch, S.~Buytaert, C.~Enz, E.~Heijne, et~al., {\it {Gasplex:
  A Low noise analog signal processor for readout of gaseous detectors}}, .

\bibitem{StripsDAQCASTclasic}
T.~Geralis, G.~Fanourakis, Y.~Giomataris, and K.~Zachariadou, ``{The data
  acquisition of the Micromegas detector for the CAST experiment}.''
\newblock IEEE Nucl. Sci. Symp. Conf. (2003) , 5 ,3455-99.

\bibitem{MATACQ}
D.~Breton, E.~Delagnes, and M.~Houry, ``{Very high dynamic range and high
  sampling rate VME digitizing boards for physics experiments}.''
\newblock IEEE TRANSACTIONS ON NUCLEAR SCIENCE, VOL. 52, NO. 6, DECEMBER 2005.

\bibitem{Iguaz:2011xj}
F.~Iguaz, T.~Dafni, E.~Ferrer-Ribas, J.~Galan, J.~Garcia, et~al., {\it {The
  Discrimination capabilities of Micromegas detectors at low energy}},  {\em
  Phys.Procedia} {\bf 37} (2012) 1079--1086,
  [\href{http://xxx.lanl.gov/abs/1110.2643}{{\tt arXiv:1110.2643}}].

\bibitem{Agostinelli:2002hh}
{\bf GEANT4} Collaboration, S.~Agostinelli et~al., {\it {GEANT4: A Simulation
  toolkit}},  {\em Nucl.Instrum.Meth.} {\bf A506} (2003) 250--303.

\bibitem{alfredotesis}
A.~Tom\'as, {\em Development of Time Projection Chambers with Micromegas for
  Rare Event Searches}.
\newblock PhD thesis, Universidad de Zaragoza, Zaragoza (Spain), 2013.
\newblock \emph{JINST} TH 001.

\bibitem{Biagi:1999nwa}
S.~Biagi, {\it {Monte Carlo simulation of electron drift and diffusion in
  counting gases under the influence of electric and magnetic fields}},  {\em
  Nucl.Instrum.Meth.} {\bf A421} (1999), no.~1-2 234--240.

\bibitem{Luzon:2006sh}
G.~Luzon, J.~Carmona, S.~Cebrian, F.~Iguaz, I.~G. Irastorza, et~al., {\it
  {Characterization of the Canfranc underground laboratory: Status and future
  plans}}, .

\bibitem{alphaguard}
\texttt{http://www.genitron.de/products/alpha\_slides.html}.

\bibitem{Thomas_SegmentedMesh}
T.~Papaevangelou, ``{Devolopment of Micromegas detectors for neutron
  time-of-flight measurements}.''
\newblock Talk given at the 3rd International Conference on Micro Pattern
  Gaseous Detectors, MPGD2013, Zaragoza (Spain), 1-4 July 2013.

\bibitem{AFTER}
P.~Baron, D.~Besin, D.~Calvet, C.~Coquelet, X.~D.~L. Broise, E.~Delagnes,
  F.~Druillole, A.~L. Coguie, E.~Monmarthe, and E.~Zonca, {\it {AFTER, an ASIC
  for the Readout of the Large T2K Time Projection Chambers }},  2008.

\bibitem{19273929}
P.~Baron, D.~Besin, D.~Calvet, C.~Coquelet, X.~D.~L. Broise, E.~Delagnes,
  F.~Druillole, A.~L. Coguie, E.~Monmarthe, and E.~Zonca, {\it {Architecture
  and Implementation of the Front-End Electronics of the Time Projection
  Chambers in the T2K Experiment}},  {\em IEEE Transactions on Nuclear Science}
  {\bf 57} (2010) 406--411.

\bibitem{AGET}
P.~Baron et~al.
\newblock Nuclear Science Symposium Conference Record, 2011.

\end{thebibliography}

\end{document}